\documentclass[aps,twocolumn,amsmath]{revtex4}

\usepackage{graphicx}

\def\diff{\mathop{\rm\mathstrut d\!}\nolimits}
\def\C60{C$_{60}$}

\begin{document} 

\title{Phase and glass transitions in short-range central \\
potential model systems: the case of \C60} 
\author{Maria C. Abramo, Carlo Caccamo%
\footnote{Email: {\tt carlo.caccamo@unime.it}},
Dino Costa, and Romina Ruberto}
\affiliation{Dipartimento di Fisica, Universit\`a degli Studi di Messina \\
Contrada Papardo, C.P. 50 -- 98166 Messina, Italy}

\begin{abstract}
 Extensive molecular dynamics simulations
show that a short-range central potential,
suited to model \C60, undergoes a high temperature transition 
to a glassy phase characterized
by the positional disorder of the constituent particles.
Crystallization, melting and sublimation, which also take 
place during the simulation runs, are illustrated in detail.
It turns out that vitrification and the mentioned phase transitions 
occur when the packing fraction of the system~---~defined in terms of 
an effective hard-core diameter~---~equals that of 
hard spheres at their own glass and melting transition,
respectively.
A close analogy also emerges between our findings
and recent mode coupling theory calculations of
structural arrest lines in a similar model
of protein solutions. We argue that the conclusions of the 
present study might hold for a wide class of potentials 
currently employed to mimic interactions in complex fluids
(some of which of biological interest), suggesting how to achieve
at least qualitative predictions of vitrification and crystallization
in those systems.
\end{abstract}

\maketitle

\columnseprule 0pt

\section{Introduction}

  Central, short-range potential models of simple analytic form 
have been the object of intense investigation over the
last years. The main reason of such an interest is that they 
provide an approximate representation of the effective interactions
and phase behavior
experimentally observed in a variety of macrosized particle systems, 
spanning from colloidal suspensions and protein or polymer
solutions \cite{zamora,lomakin,piazza2,
poon1,tenwolde,louis,foffi,pellicane}
 to other macromolecular systems and
fullerenes \cite{ cheng,hagen,ashcroft,hasegawa,
costa,costa1}.   
A paradigmatic example
in this respect is offered by 
protein crystallization~\cite{chayen,mcpherson,broide}:
indeed,
if one adopts a representation in which the macromolecules 
are spherical particles 
interacting through a central short-range potential, 
and the solvent is assumed to be 
a structureless continuum,
the liquid-vapor binodal and the sublimation line thereby
calculated exhibit all features of the
protein-rich/protein-poor and 
 crystallization line observed
in real globular protein solutions.
These models then offers an unvaluable tool for the study,
by usual means of statistical mechanics and simulation,
of the relative location of phase coexistence lines
in real systems.  

  This scenario 
has been further enriched by recent investigations of
glass transition in one of the most studied  
short-range model,
namely the attractive hard-core Yukawa fluid. 
In this context the mode coupling theory
predicts the onset of two glassy phases
deeply inside the metastable liquid-solid region of the model~\cite{foffi}. 
The formation 
of distinct vitreous phases with peculiar internal characteristics has also
been reported in recent experiments on colloid-polymer
 mixtures~\cite{bartsch,chen,poon}.
   
These evidences
prompted us to an extensive investigation of
phase coexistence and glass transition conditions
 in short-range models, with particular attention to the rich
phenomenology which characterizes the 
metastable region enveloped between the freezing
and the melting line of such systems. 
  The paper 
focuses in particular on 
the Girifalco potential for fullerene \C60~\cite{girifalco}. 
This model can be considered as ``marginally'' short-range 
in the sense that, whereas in colloidal and
protein systems 
the interaction between (model) macroparticles
 reduces practically to zero 
within a fraction of the particle diameter
and the liquid-vapor equilibrium is only metastable, 
in the Girifalco model 
the decay length of interactions is slightly greater than the
fullerene diameter and a stable liquid phase does exist, 
albeit confined to a temperature interval smaller
than 100\,K~\cite{cheng,hasegawa,costa,costa1}, with
no experimental evidence of it hitherto 
reported~\cite{fischer,xu,sundar,noi,nota}. 
Such characteristics appear consistent with early  
speculations about the ``elusive diffusive'' 
nature of liquid \C60~\cite{ashcroft},
as well as with recent evidences of the absence, in this peculiar
fluid, of cage effects in the velocity autocorrelation functions,
and of collective phenomena, ordinarily 
observed in simple liquids~\cite{alemany}.

We have recently reported in 
a preliminary communication~\cite{vetro} (hereafter referred to as I)
Molecular Dynamics (MD) evidences of the 
onset of a glass transition in \C60 at $T=1100$\,K upon a 
pressure of 3.5\,MPa, as associated to positional
disorder. Such a glassy phase is quite distinct 
from the experimentally detected \C60 orientational glass, 
which forms at 90\,K as a consequence of the freezing at 
low temperature of the residual disorder in the orientation
 of the fullerene cages~\cite{gugenberger,matsuo}.

  As preliminary documented in I, vitrification of \C60 at high temperature
is assessed through quenching cycles of
 the liquid phase at various pressures, by 
analyzing a number of thermodynamic, structural and dynamical quantities.
  Here, we report and discuss in detail the MD results 
we cumulated to support our conclusion that a positional glass
of \C60 is effectively formed. In the same context,
we discuss other transitions undergone by the 
system in the course of cooling or heating cycles,
such as crystallization, melting and sublimation.
As we shall detail, such transitions
take place deeply inside the metastable region in coincidence of
the crossing of two density-vs-temperature {\it loci}
over which the ``effective'' packing fraction
of the system equals that of hard spheres at their melting
and glass transition, respectively.
   The analysis of such hard-sphere-like behavior
enlights the analogies of our study with the case, discussed
 in reference~\cite{foffi}, of the onset of glassy phases in 
model protein solutions, 
suggesting that our results
can be useful in the more general context of
complex fluids investigation.

The paper is structured as follows: in section~II we introduce the model 
and describe the computational strategies. In~III  results
for thermodynamics, structural and dynamical properties are reported. 
The discussion and conclusions follow in section~IV.

\section{Model and simulation strategies} 

\indent We study a system composed of particles 
interacting via the Girifalco potential~\cite{girifalco},
\begin{eqnarray}\label{eq:pot}
v(r)  &=&    -\alpha_1 \left[ \frac{1}{s(s-1)^3} +
                   \frac{1}{s(s+1)^3} - \frac{2}{s^4} \right]  \nonumber\\[4pt]
    & & \quad   +\alpha_2    \left[ \frac{1}{s(s-1)^9} +
                   \frac{1}{s(s+1)^9} - \frac{2}{s^{10}} \right] 
\end{eqnarray}
where $s=r/d$, $\alpha_1=N^2A/12d^6$, and
$\alpha_2=N^2B/90d^{12}$;
$N=60$ and $d=0.71$\,nm are the number of carbon atoms and the diameter,
respectively, of the fullerene particles,
$A=32\times10^{-60}$\,erg\,cm$^6$ and
$B=55.77\times10^{-105}$\,erg\,cm$^{12}$ are constants entering the
12-6 potential $\phi$(r)$= -A/r^6+ B/r^{12}$ through which two carbon sites on
different spherical molecules are assumed to interact.
The distance where the potential~(\ref{eq:pot}) crosses zero,
the position of the potential well minimum and its depth, 
are $\sigma=0.959$\,nm, $r_{\rm min}= 1.005$\,nm, and 
$\varepsilon = 0.444\times10^{-12}$\,erg, respectively~\cite{girifalco}. 
We assume $v(r)=\infty$ for $s<1$.

We have used
the MD method introduced by Andersen~\cite{andersen}
to simulate constant-pressure, constant-enthalpy (NPH) conditions,
and the Verlet algorithm to integrate the equations of motion
over time steps $\Delta t=5$\,fs;
cubic boxes and periodic boundary conditions are assumed.
Most results are obtained for $N=864$ and $N=1000$
total number of particles; the smaller sample is fully compatible with
an initial fcc lattice arrangement of the particles in a cubic simulation
box, whereas the 1000-particle distribution obviously 
implies defects.
Both sizes are part of a more general analysis
employing also five hundred and 1372 particles,
to exclude any dependence 
of simulation results on the sample size. 
  A wide range of pressures, spanning from close-to-atmospheric conditions
up to 250\,MPa (2.5\,kbar), has been explored 
to verify the effects on crystallization and glass transition
conditions.  Cooling and quenching cycles 
start from an initial supercritical or liquid  
configuration (see figure~\ref{fig:1}), identified as such
on the basis of the known phase diagram of
\C60~\cite{costa1}.
Heating cycles begin either from the defective crystals
eventually obtained at the end of the cooling
cycles, or from a perfect fcc crystal at room temperature 
with 864 particles. 
Other details are given in the presentation of 
results.

\section{Results}

For clarity sake, we shall detail in the first instance the  
transitions undergone by the system
during slow cooling or heating 
sequences, followed by a presentation of results obtained
during the quenching routines.
Just before, we
introduce the concept of ``effective'' packing
fraction for the system at issue, with two associated loci 
of thermodynamic states.

\subsection{Effective hard sphere diameters and  \\
constant ``packing'' loci}

We introduce
an effective hard sphere diameter for 
the \C60-\C60 interaction with the 
definition of a reference potential for $v(r)$
of equation~(\ref{eq:pot})
according to the well-known Weeks-Chandler-Andersen (WCA)
prescription~\cite{weeks}, 

\begin{eqnarray}\label{eq:wca}
\begin{array}{cc}
v_{\rm ref}(r) = &
\left\{
 \begin{array}{cc}
 v(r)+\varepsilon    & \quad {\rm if} \ r \le r_{\rm min} \cr
 0                      & \quad {\rm if} \ r > r_{\rm min}
 \end{array}
 \right. \\ \\
\end{array} \,,
\end{eqnarray}
and adopt the Barker and Henderson
 expression for the effective hard-core diameter~\cite{BH},
\begin{equation}\label{eq:sigma1}
\sigma_{\rm BH}=\int_0^\infty \{1 -\exp[-\beta v_{\rm ref}(r)]\}dr \,.
\end{equation}
The Barker-Henderson prescription for the hard sphere
diameter is by no means unique. Other more sophisticated recipes,
hinging for instance on the use
of the cavity distribution 
function $y(r)=\exp[\beta v(r)]g(r)$~\cite{report}
(where $g(r)$ is the radial distribution function),
have been proposed in the literature. 
We have verified however that the resulting changes in the 
estimate of the effective diameter 
with respect to the simpler BH prescription are quite minor.

We now look for densities $\rho=\rho(T)$ 
which make the effective  
packing fraction $\eta=\pi/6\rho\sigma_{\rm BH}^3$ 
constant with respect to temperature
variations, and equal to a prefixed value.
 In particular, we impose $\eta$ to be equal 
to the packing of hard spheres at their own 
melting and glass transition,
respectively.
The value  $\eta^{\rm HS}_{\rm m}=0.545$
 has been reported for the packing fraction
of hard spheres at
melting \cite{hoover}. A  packing of 0.55 has also been 
reported in constant pressure MD simulation
of hard spheres \cite{gruhn}. Here we choose an intermediate value
(see I), by imposing
\begin{equation}\label{eq:etamelt}
\eta_{\rm m}=\pi/6\rho_{\rm m}(T)\sigma_{\rm BH}^3=0.548\,.
\end{equation}
As for the packing of hard
spheres at the glass transition, $\eta_{\rm g}^{\rm HS}$,
the value $0.58$ has been reported in reference~\cite{woodcock}. 
whereas effective packing fractions at vitrification
for continous potentials, mostly of Lennard-Jones-like form
(see, e.g.~\cite{shumway} and references),
approach 0.58
with a considerable dispersion of values
due to the 
different prescriptions either for the effective diameter
or for the determination of the glass transition
temperature.  We require in particular (see I):  
\begin{equation}\label{eq:etaglass}
\eta_{\rm g}=\pi/6\rho_{\rm g}(T)\sigma_{\rm BH}^3=0.574\,.
\end{equation}
On the basis of equations~(\ref{eq:etamelt}) 
and~(\ref{eq:etaglass})  we can determine
the two functions $\rho_{\rm m}(T)$ and $\rho_{\rm g}(T)$
which are shown in figure~\ref{fig:1}.
As we shall see, these two $\rho(T)$ loci
correlate significantly with the crystallization and vitrification
conditions of the \C60 model.

\subsection{Cooling cycles and crystallization; heating cycles
and sublimation or melting}

 We first illustrate the case when the 
pressure $P=3.5$\,MPa.
Cooling starts from a high temperature
fluid configuration at $T=2000$\,K (see figure~\ref{fig:1}, top).
In sequential stepwise 
drops $\Delta T=30$\,K, the system
 evolves 
over 20\,000 steps (correspondiong to 100 ps) at
fixed temperature, followed
by 10\,000 cumulation steps where the system evolves freely.
The final temperature $T$,  box volume $V$,
 enthalpy $H$, 
and other thermodynamic, structural and dynamical properties
are recorded at each $\Delta T$ (see below).
Cooling is always arrested when room temperature is achieved or 
sligthly before this threshold.
Statistical uncertainties on $T$, $\rho$, and $P$  
turn out to be 0.8\,K, 0.0003\,nm$^{-3}$ and 0.1\,MPa, respectively.

\begin{figure*}
\begin{center}
\includegraphics[width=6.5cm,angle=-90]{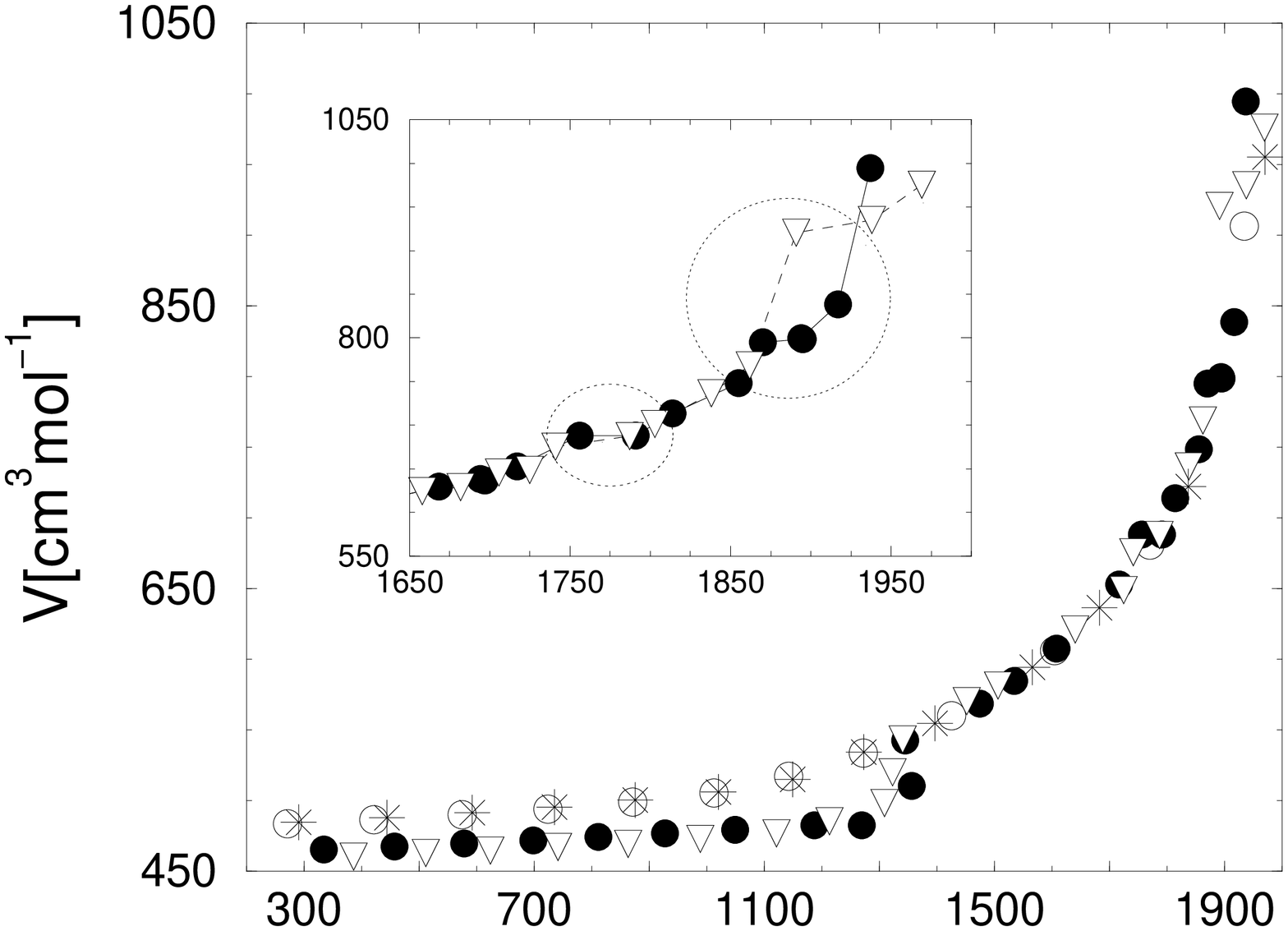} \hspace{20pt}
\includegraphics[width=6.5cm,angle=-90]{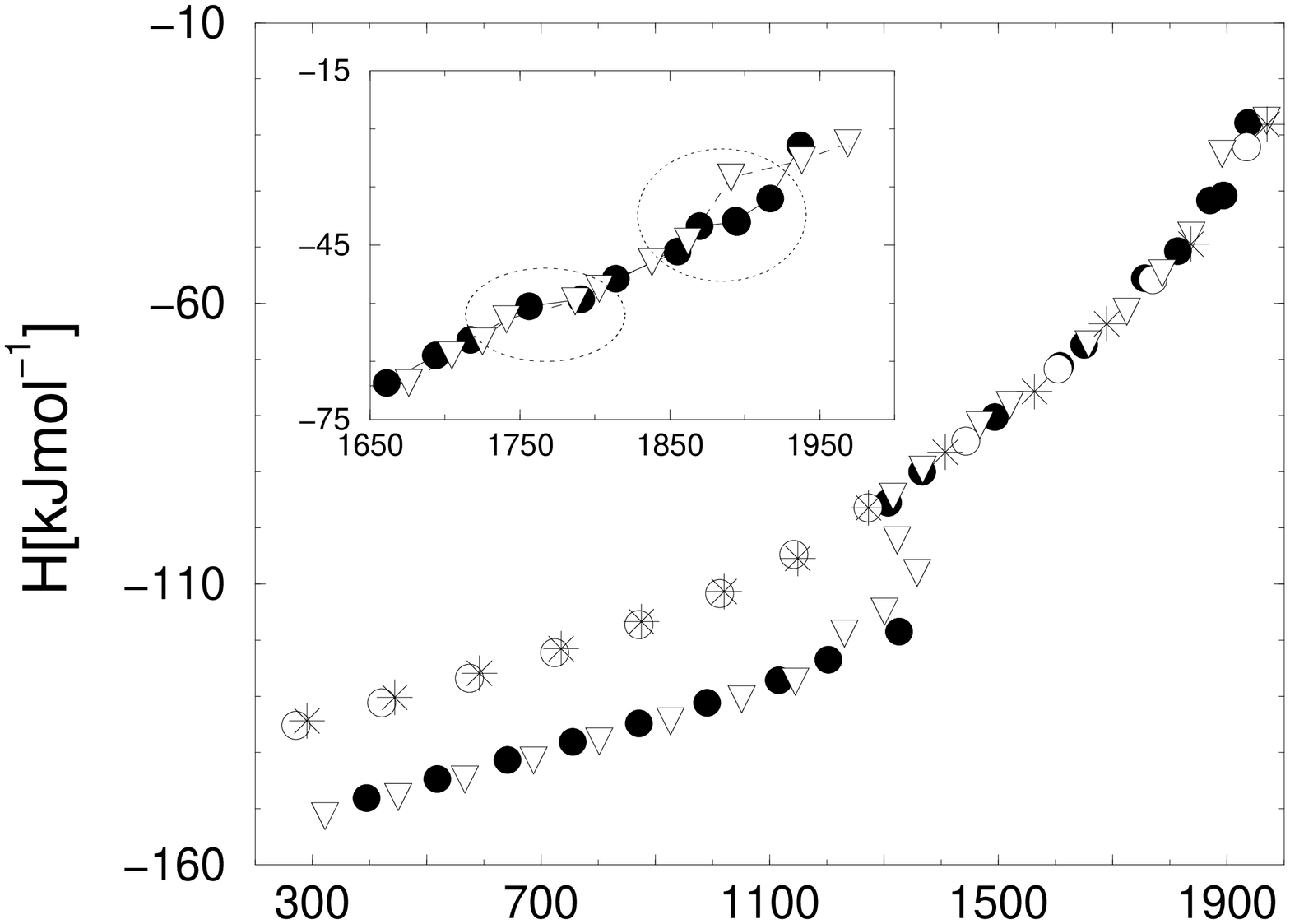}
\hspace*{10pt}\includegraphics[width=6.5cm,angle=-90]{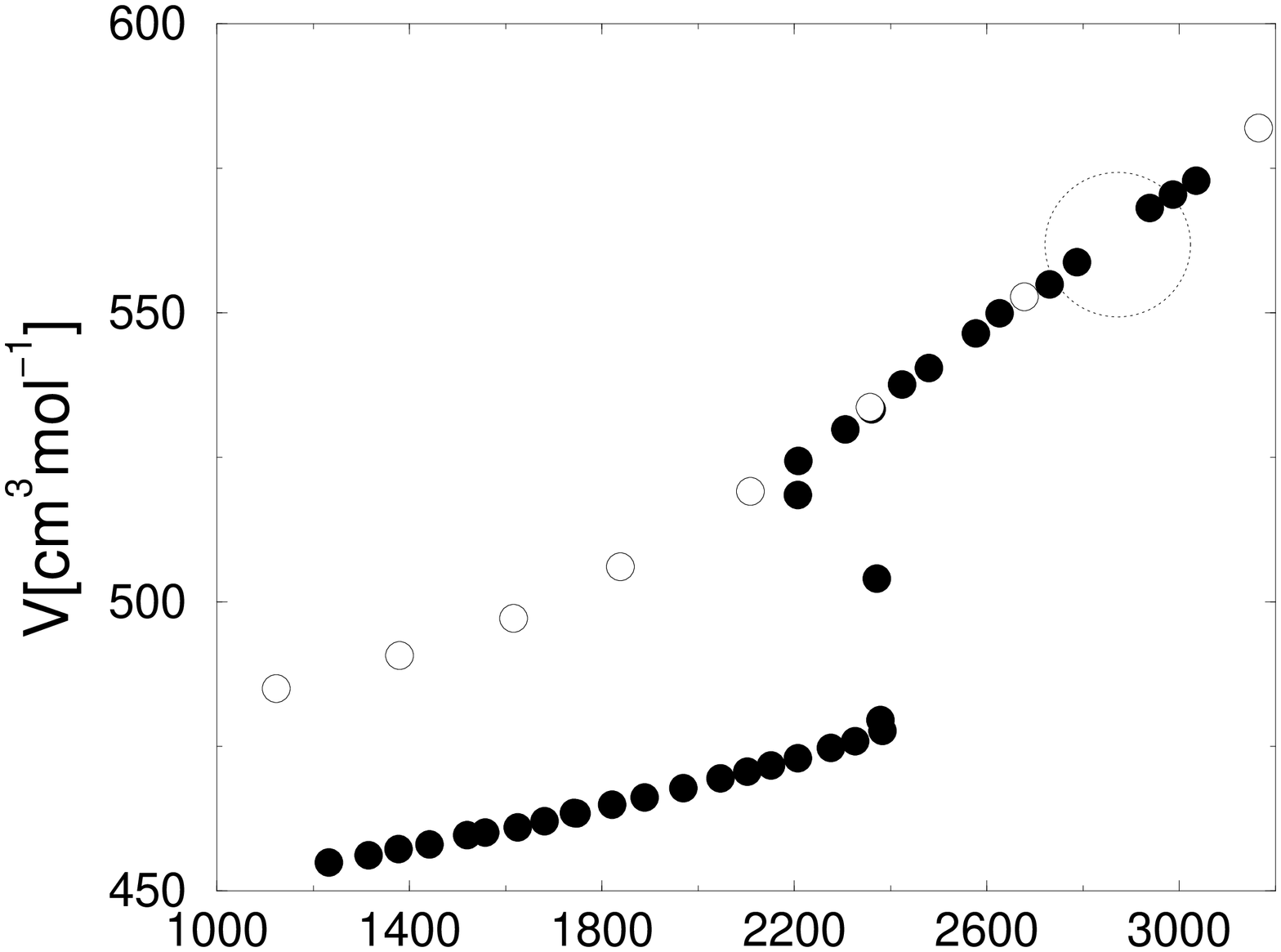} \hspace{25pt}
\includegraphics[width=6.5cm,angle=-90]{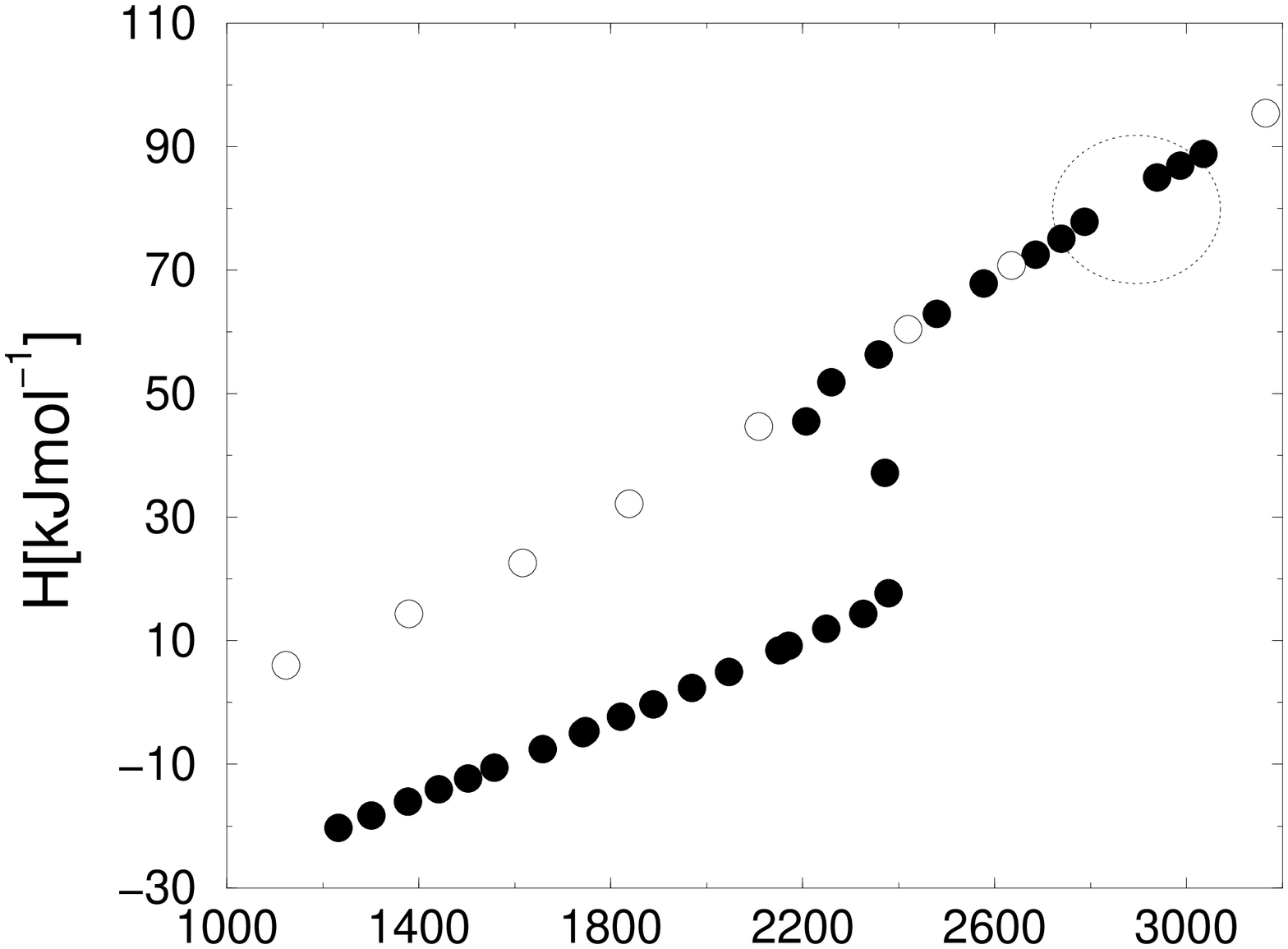}
\hspace*{15pt}\includegraphics[width=6.5cm,angle=-90]{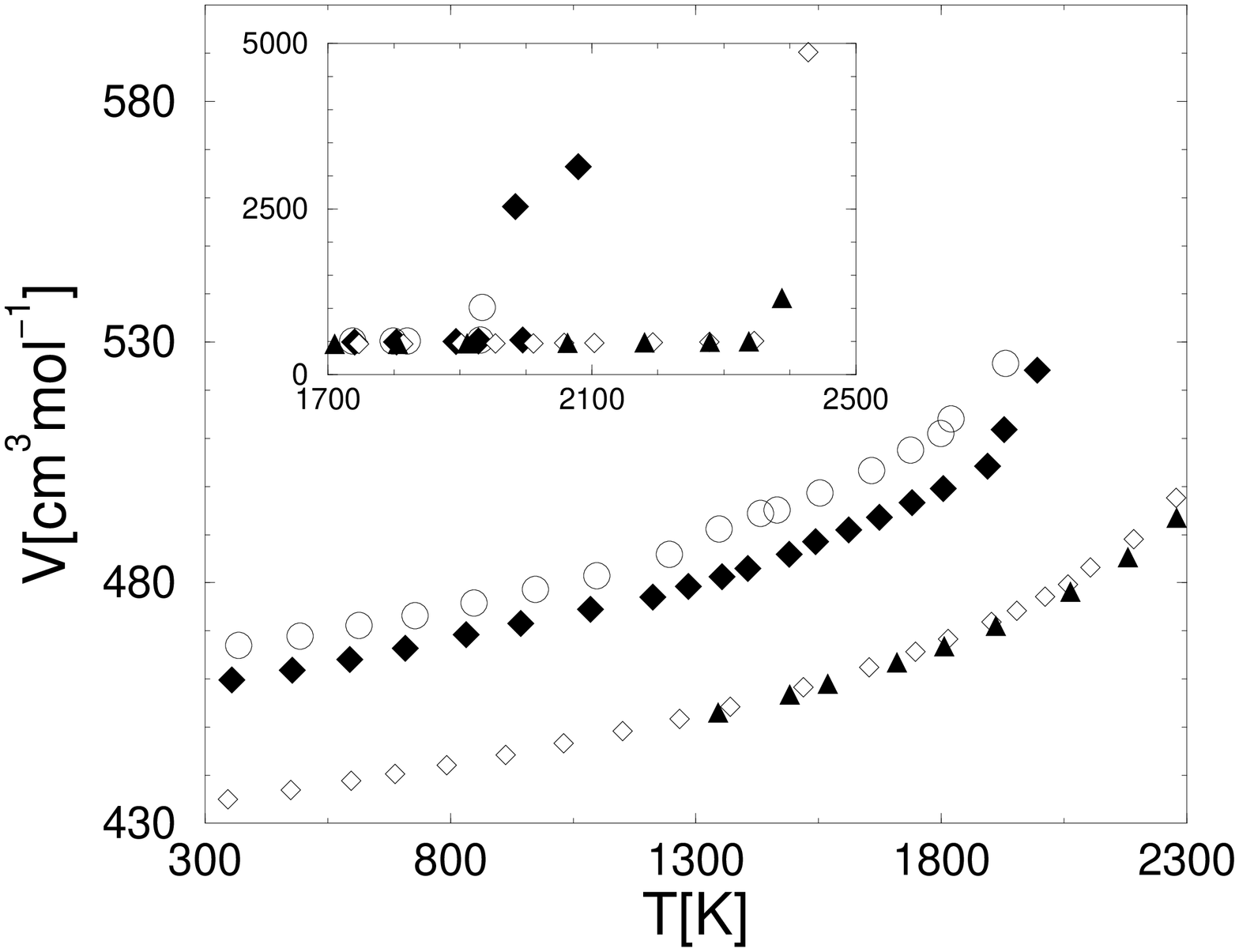} \hspace{10pt}
\includegraphics[width=6.5cm,angle=-90]{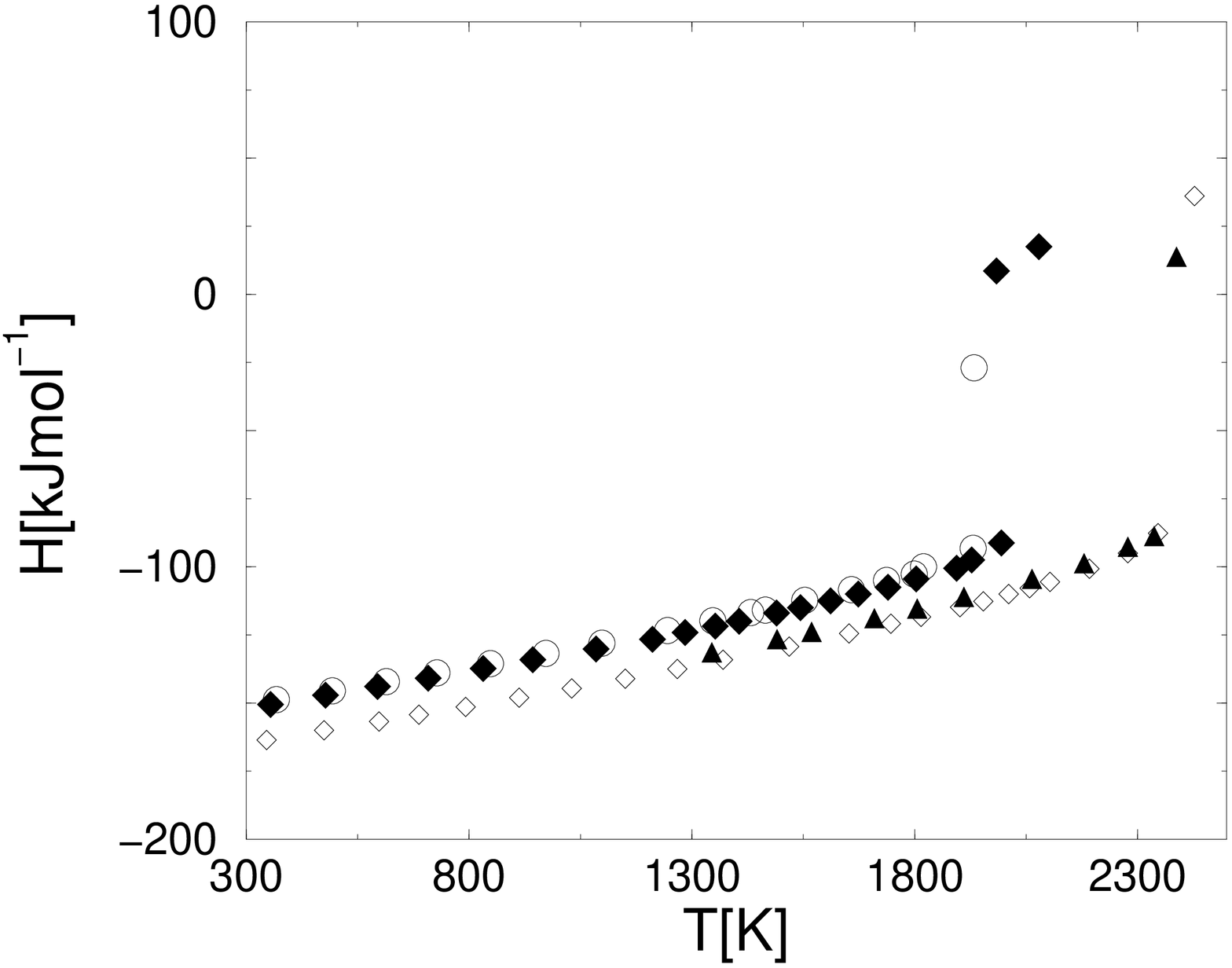}
\caption{\label{fig:2}
{\it Top:} 
Volume (left) and enthalpy (right) as functions of the  
temperature in cooling and 
quenching cycles at $P=3.5$\,MPa.
Full circles, open triangles: cooling with $N=1000$ and 864, respectively;
open circles, asterisks: quenching with $N=1000$ and 864, respectively.
Inset: expansion of the crossing zones (encircled) of the 
freezing and binodal line during cooling.
{\it Middle:} $V$ (left) and $H$ (right)
 vs $T$  in cooling (full circles) and quenching (open circles)
  cycles with $N=1000$ at  $P=250$\,MPA.
{\it Bottom:} $V$ (left) and $H$ (right) vs $T$ in heating cycles. 
 Open circles, full diamonds: heating of the defective crystal
 with $N=1000$ and 864 at $P=3.5$\,MPa, respectively;
 open diamonds, full triangles: heating of the perfect crystal
  with $N=864$ particles at $P=3.5$ and 12\,MPa,respectively.
 Inset: expansion of the volume scale around the 
 transition to the liquid (for $N=1000$ at 3.5\,MPa,
 open circles),
 and the sublimation to a highly expanded vapor phase (for all other cases).
} \end{center}
\end{figure*}

\begin{figure*}
\begin{center}
\includegraphics[width=7.8cm]{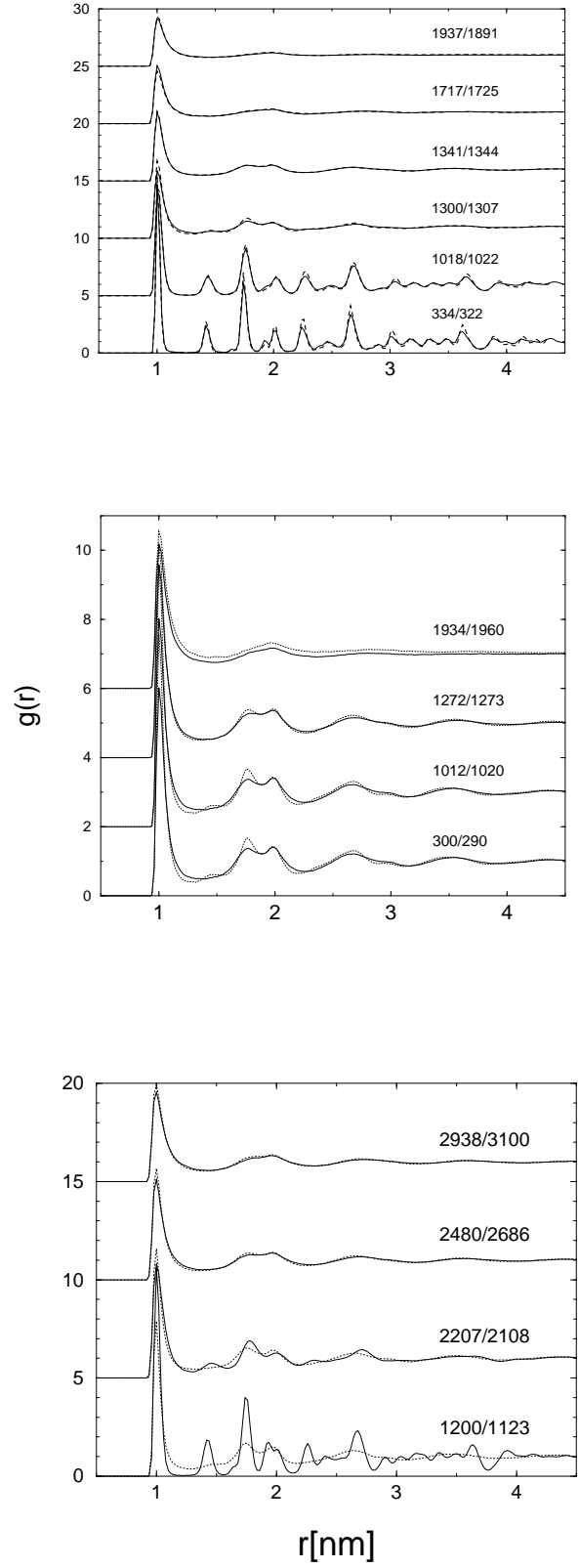}
\caption{\label{fig:3}
{\it Top:}
evolution of $g(r)$ with the temperature during cooling
at 3.5\,MPa, with $N=1000$ (full line) and $N=864$ (dashed line).
Temperatures aside the curves refer to the $N=1000$ (left) and 864 (right).
{\it Middle:}
same as in the top panel for quenching cycles.
{\it Bottom:} Cooling (full line) and quenching (dashed line) for 
$N=1000$ at $P=250$\,MPa.
}\end{center}
\end{figure*}

\begin{figure*}
\begin{center}
\includegraphics[width=7.5cm,angle=-90]{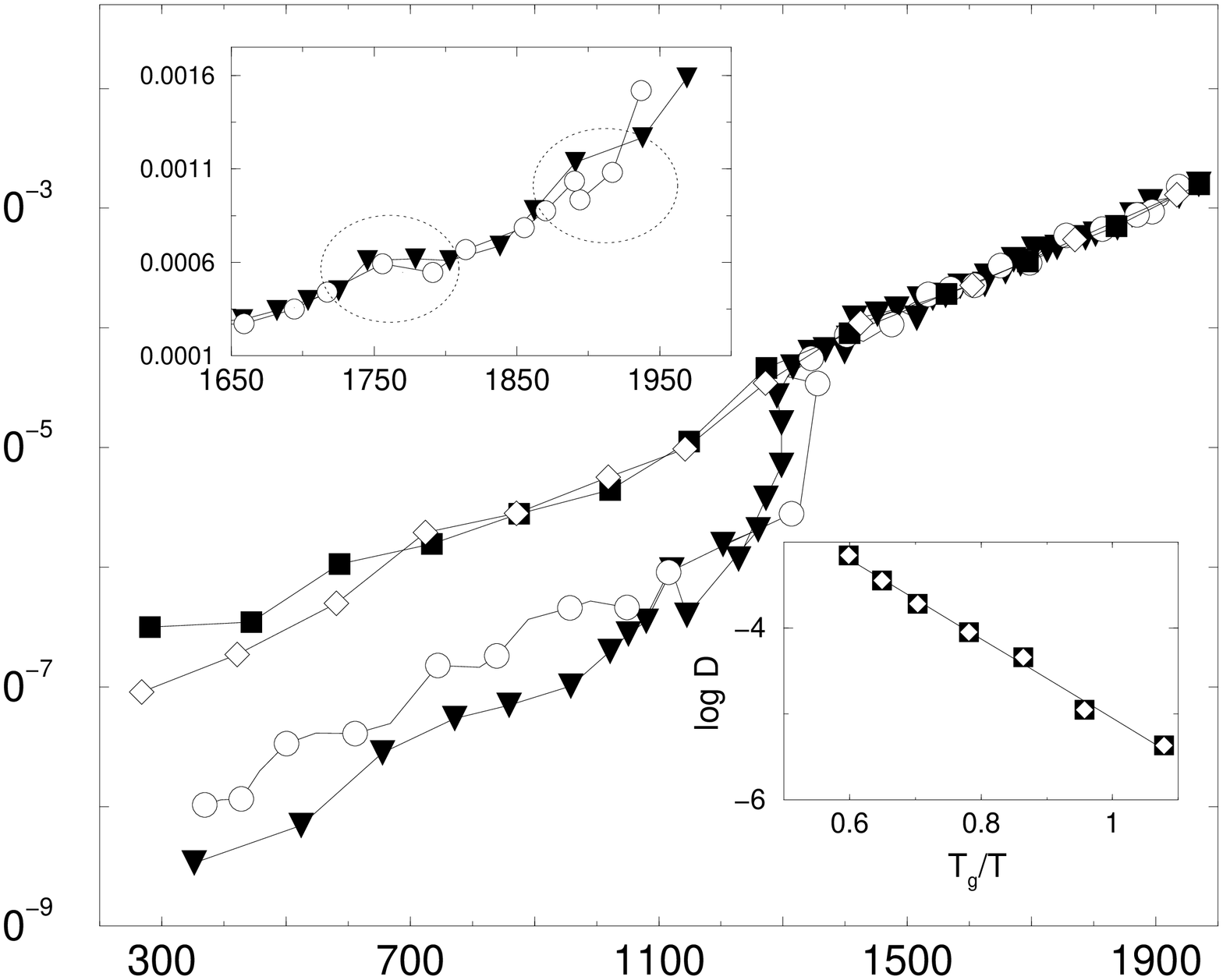} \\
\vspace{20pt}\includegraphics[width=8.4cm,angle=-90]{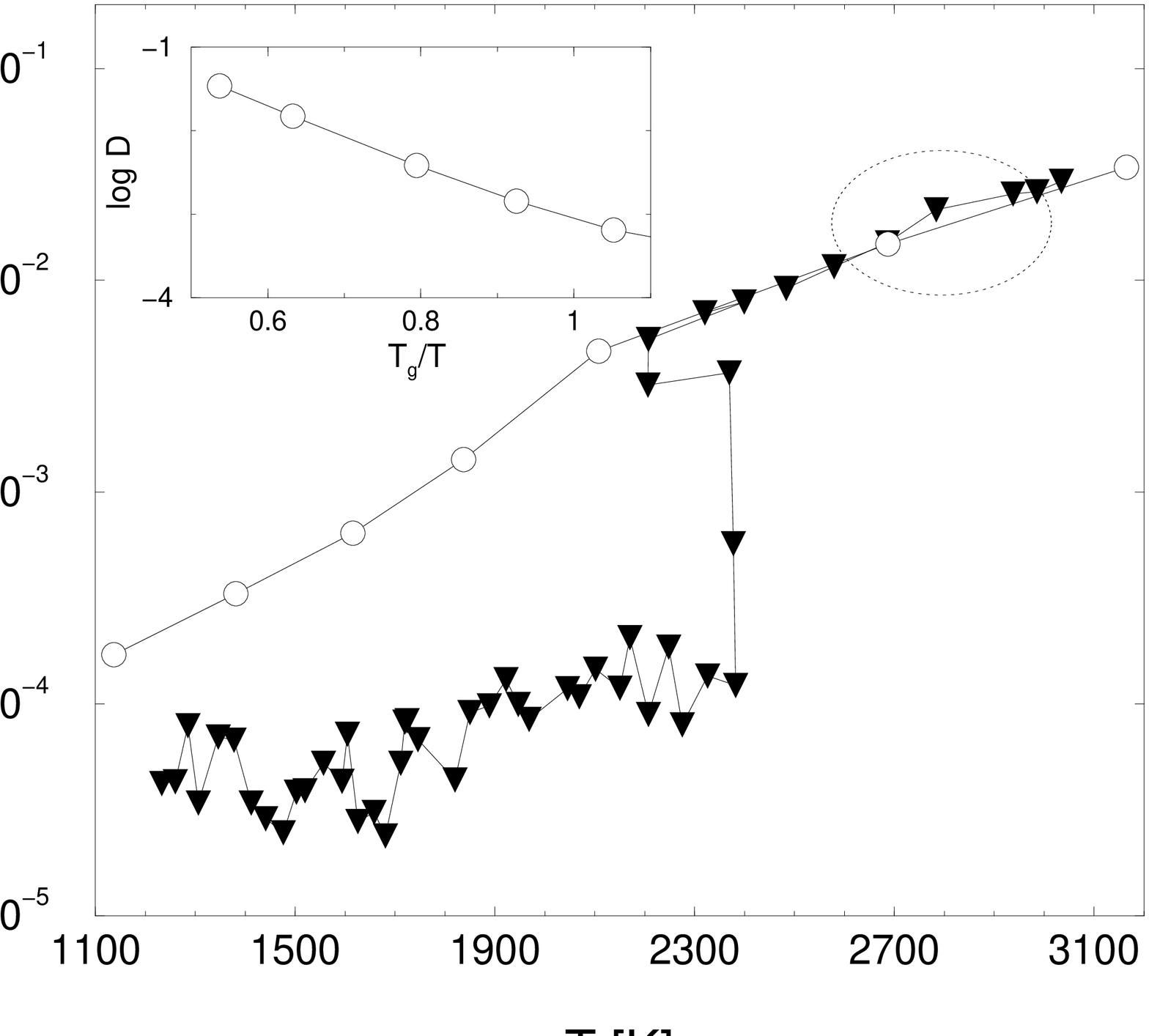} \\
\vspace{20pt}
\caption{\label{fig:4}
{\it Top:} Diffusion coefficient of cooling (open circles, full triangles) 
and quenching (open diamonds and full squares)
cycles for $N=1000$ and 864, at 3.5\,MPa, respectively. 
Upper inset: linear scale expansion of the $D$
 behaviour at the crossing of the freezing and binodal lines;
lower inset: Arrhenius plot of the diffusion coefficient in quenching cycles.
{\it Bottom}: Diffusion coefficient in cooling (full triangles) 
and quenching (open circles) cycles at $P=250$\,MPa for $N=1000$.   
Inset: Arrhenius plot of $D$ in the quenching cycles.}
\end{center}
\end{figure*}

\begin{figure*}
\includegraphics[width=4.1cm]{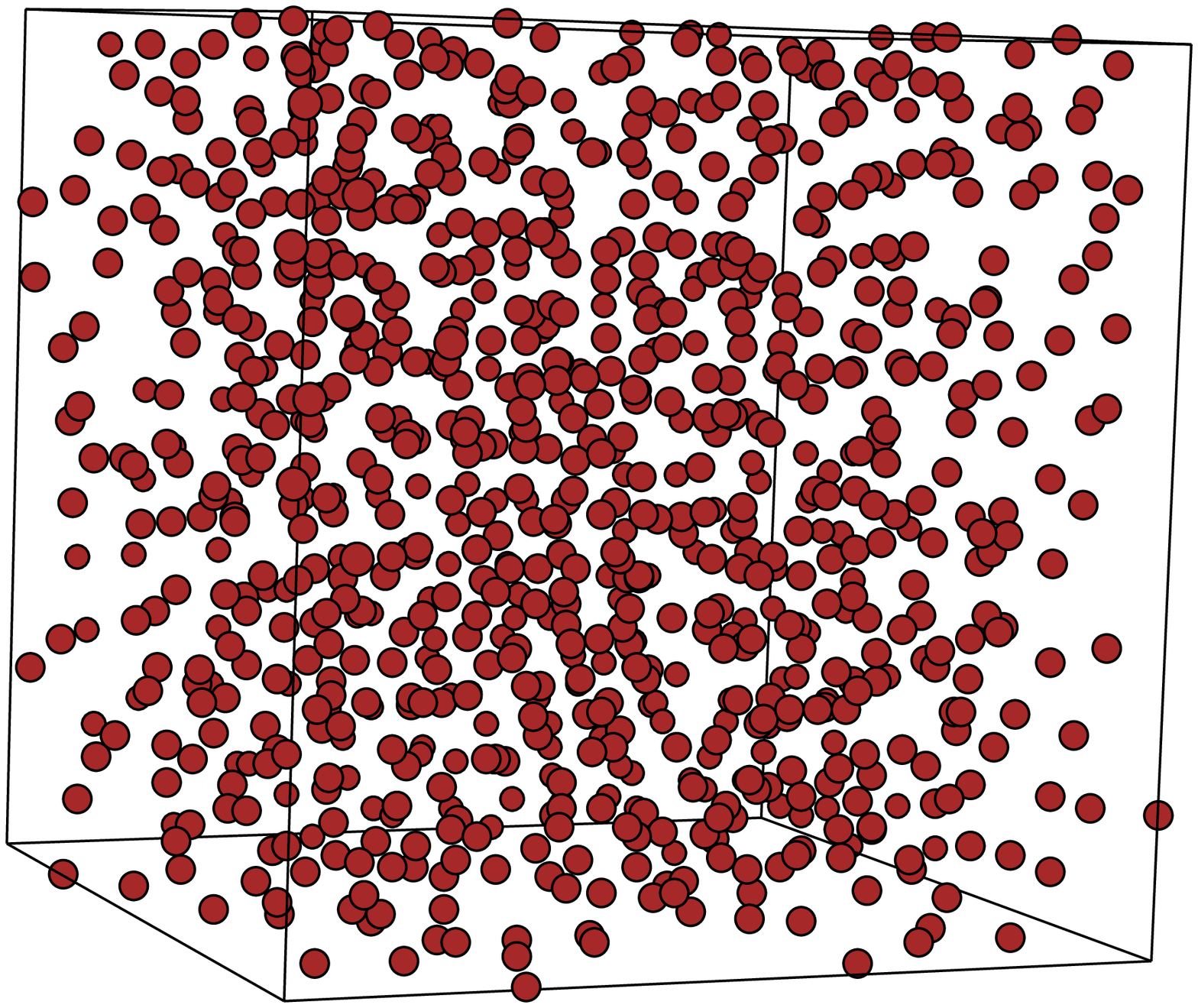}
\includegraphics[width=4.1cm]{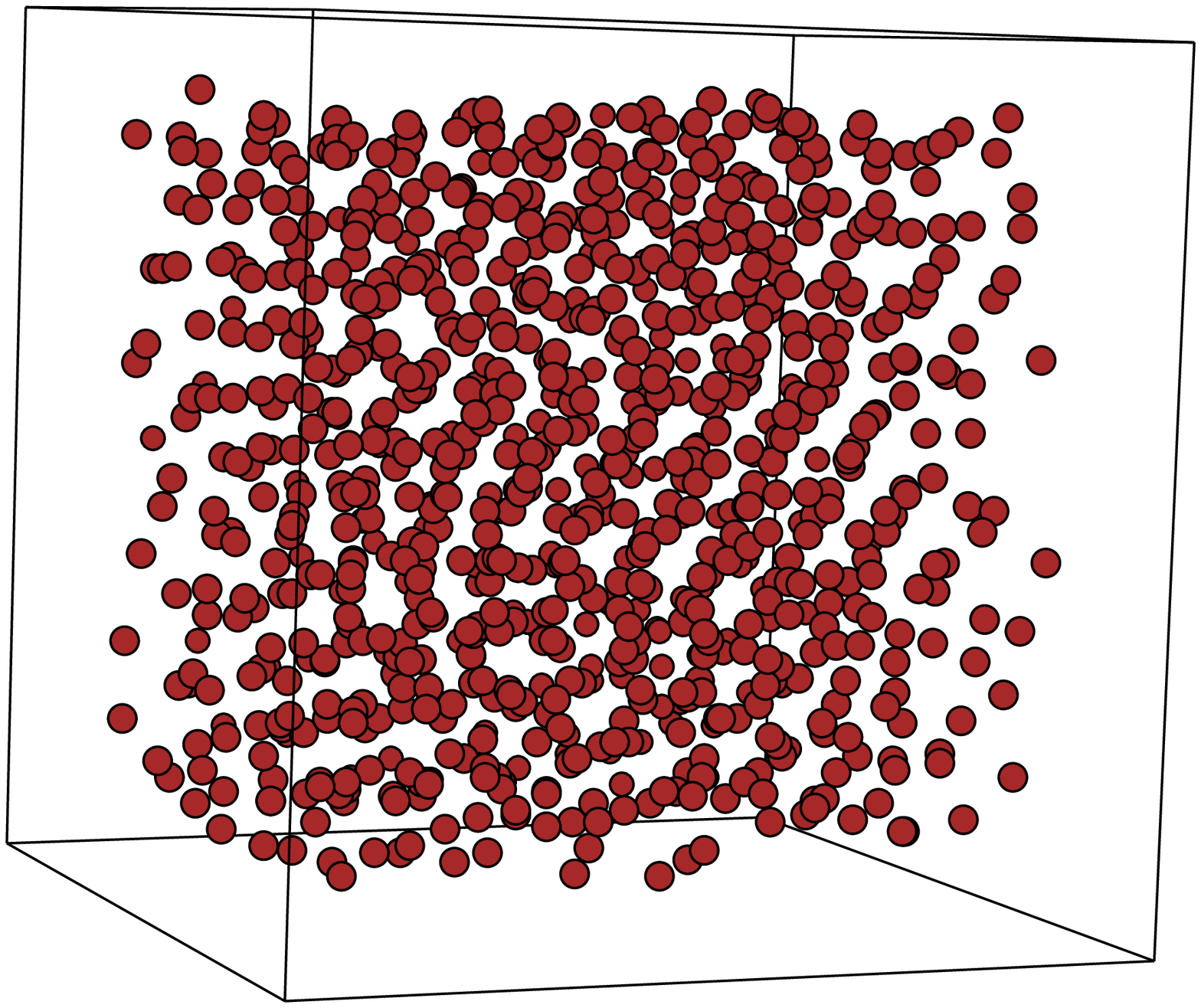}
\includegraphics[width=4.1cm]{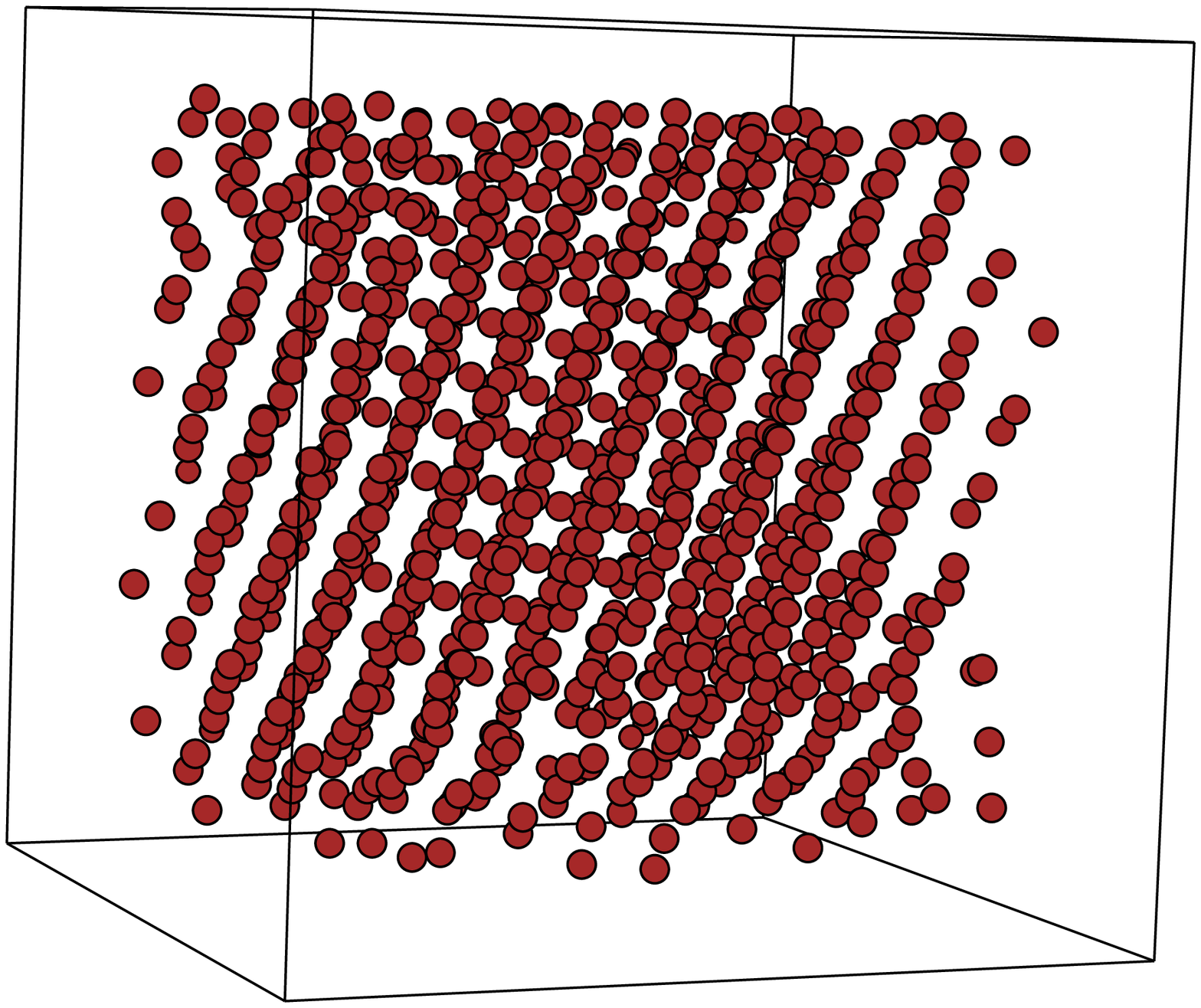}
\includegraphics[width=4.1cm]{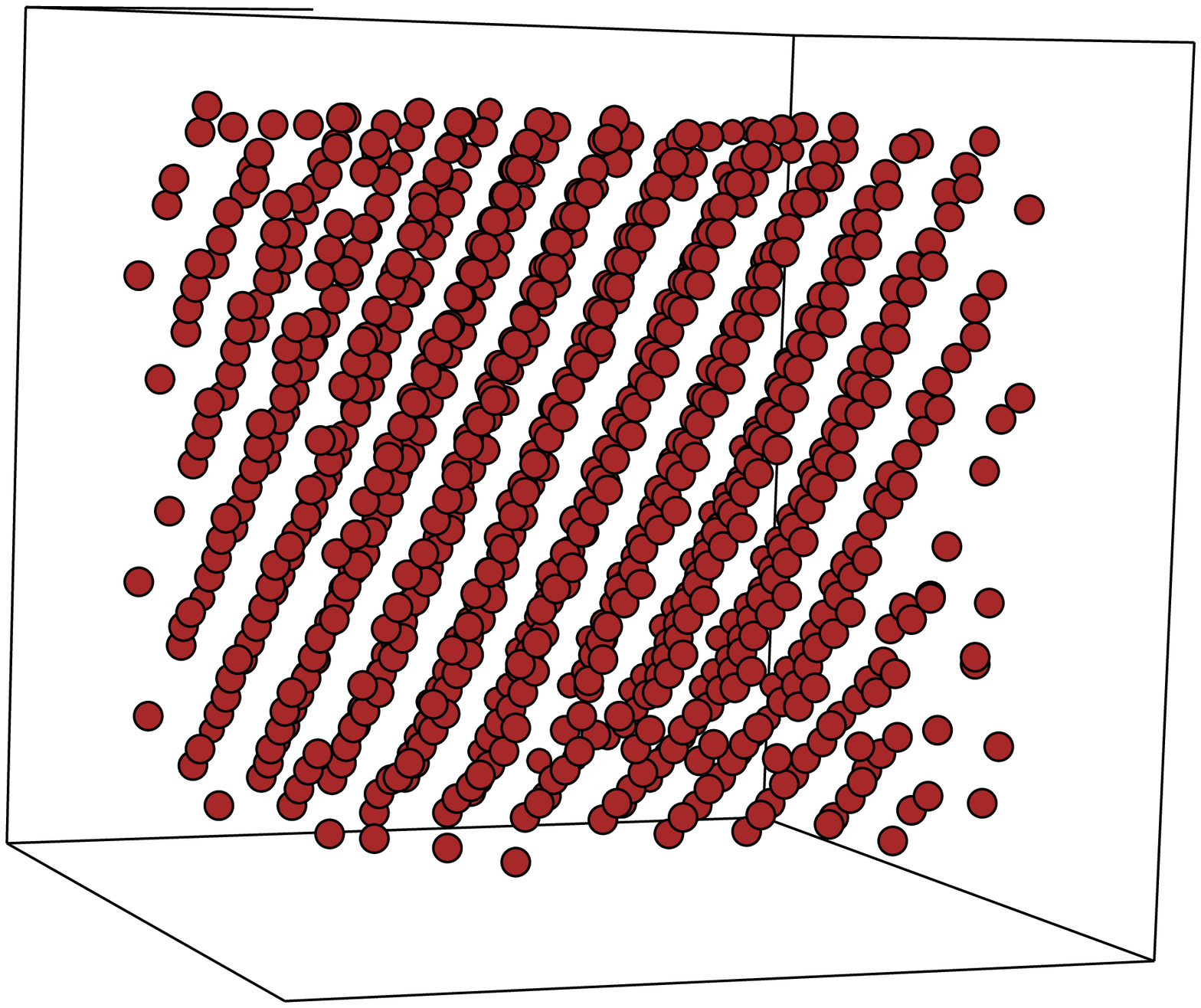}\\ 
\includegraphics[width=4.1cm]{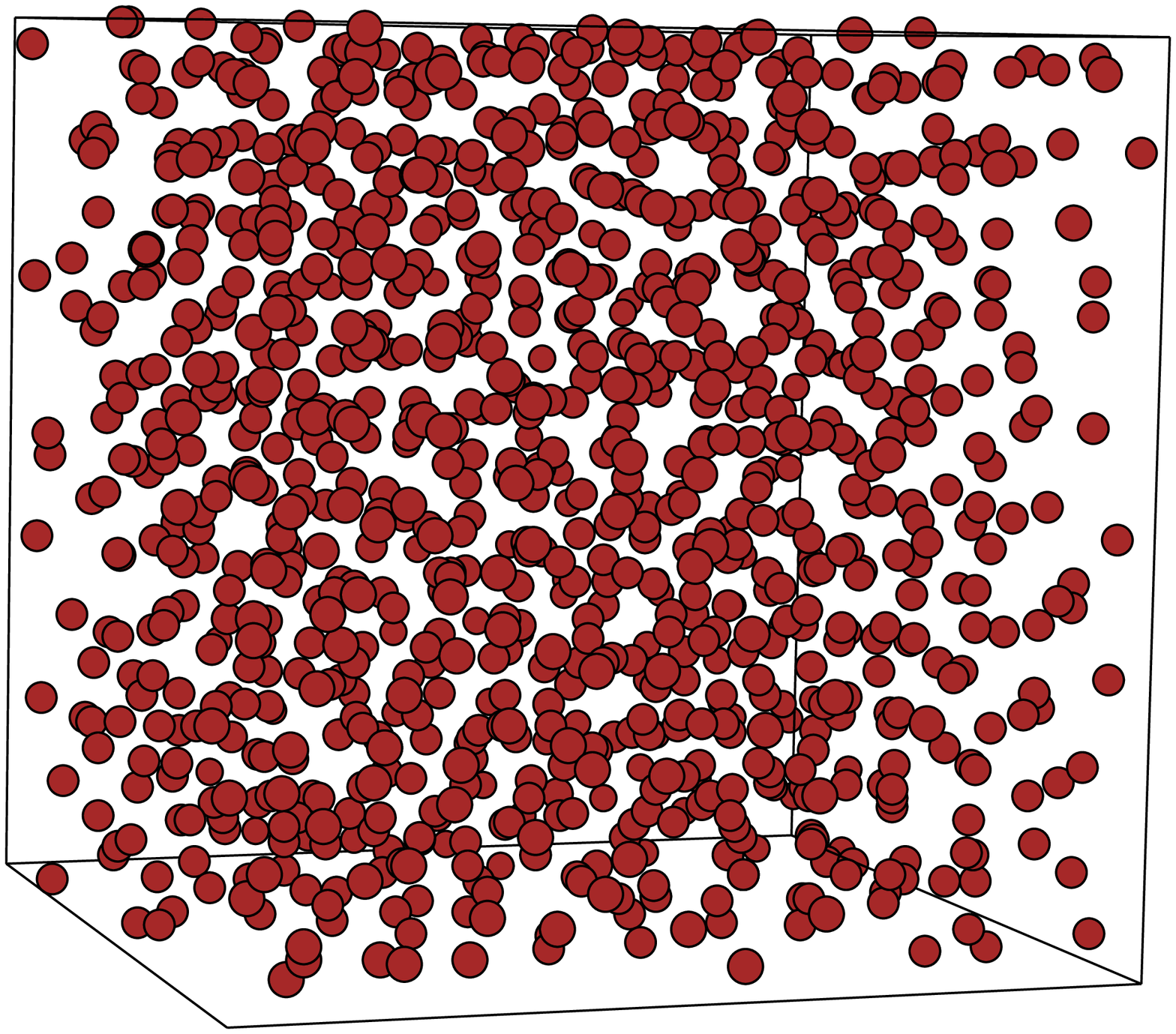}
\includegraphics[width=4.1cm]{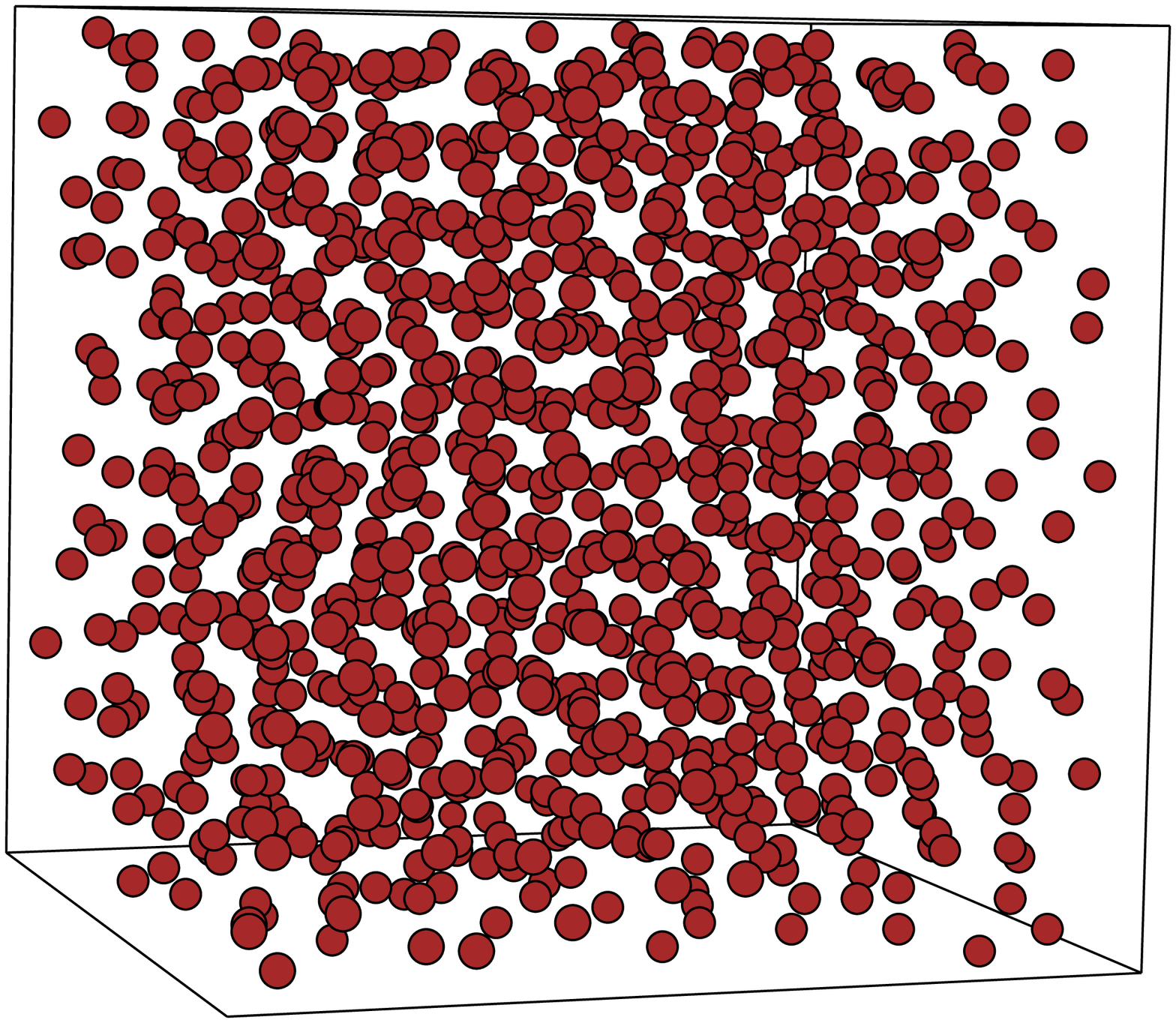}
\includegraphics[width=4.1cm]{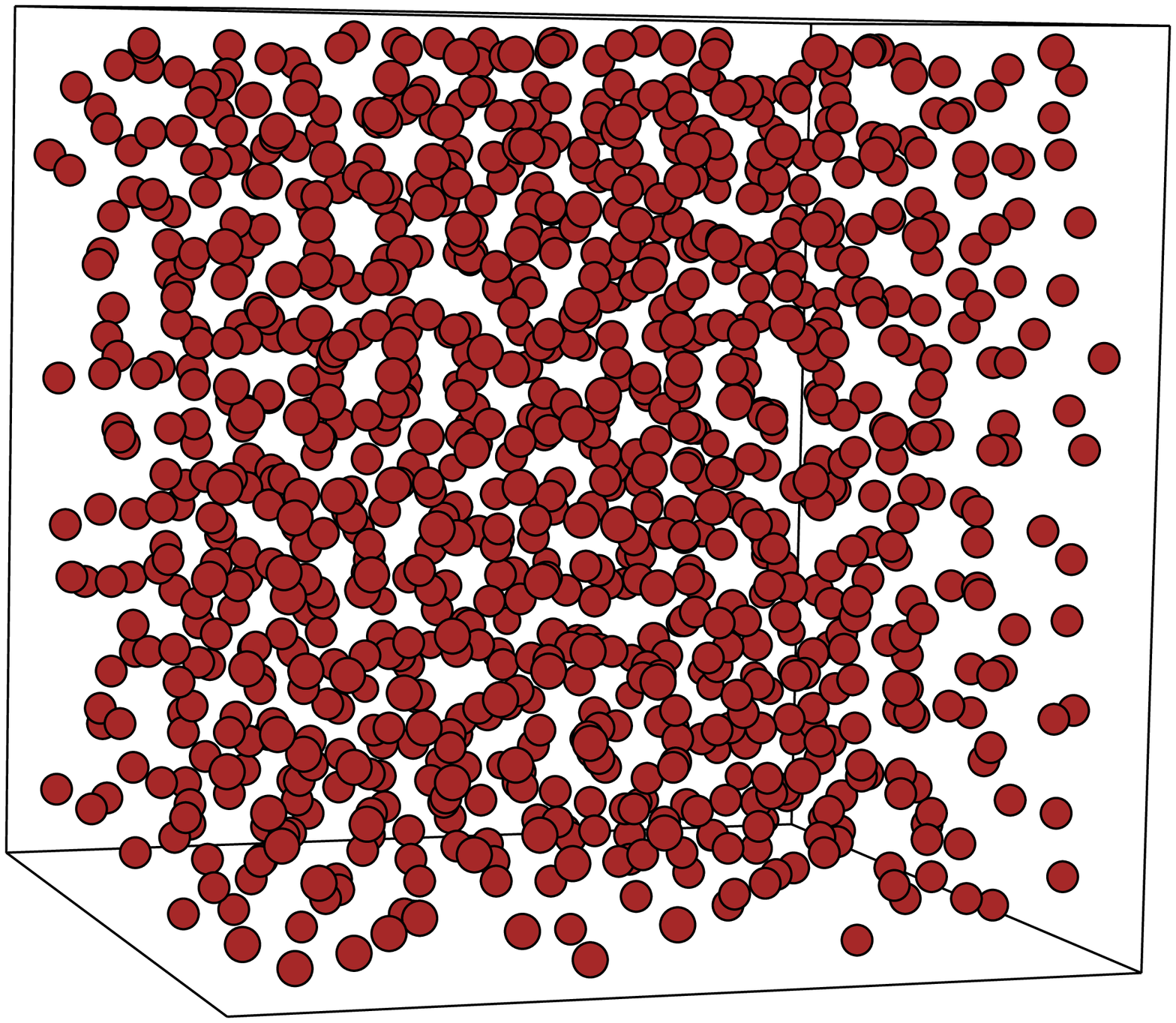}
\includegraphics[width=4.1cm]{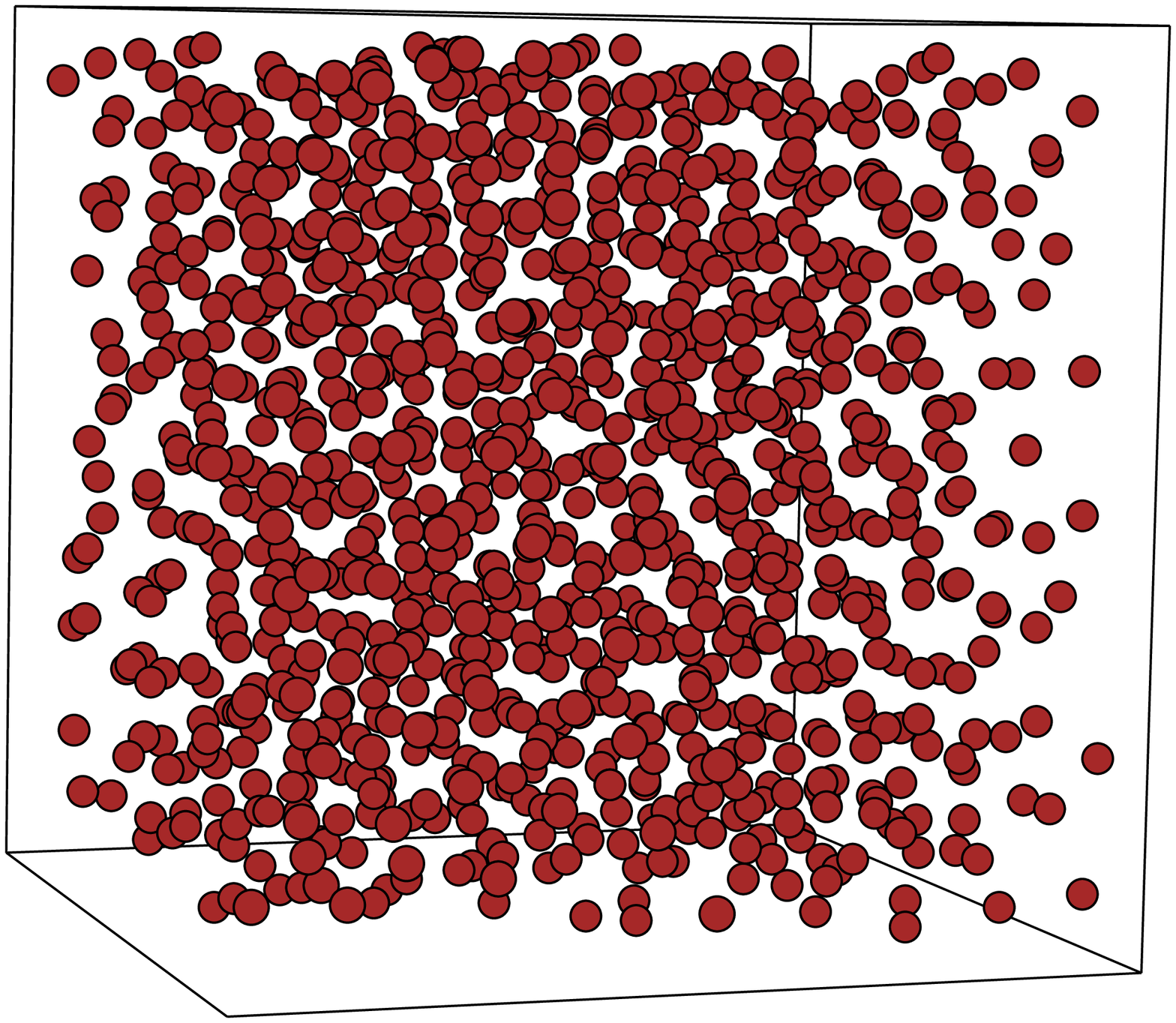}
\caption{\label{fig:5}
Snapshots of the $N=1000$ sample 
during the cooling (top) and the quenching (bottom) sequence
at $P=3.5$\,MPa and
$T=1937$, 1307, 1152, and 300\,K (from left to right).
} \end{figure*}

Cooling paths are visualized in figure~\ref{fig:1} (top)
 in the $\rho$-$T$ representation of the phase diagram of \C60. 
Volumes and enthalpies changes with the temperature are
 displayed in figure~\ref{fig:2} (top);
as visible
 (and already shown in I for the case $N=1000$),
for both system sizes investigated
$V$ and $H$ undergo a marked drop 
 at $T\simeq 1300$\,K, accompanied by
a temporary increase of the temperature.
Such a highly nonmonotonic behaviour
of the two thermodynamic quantities
is fully consistent with a transition of the system to
the solid phase
as further documented by {\it (i)} the behavior of the 
radial distribution function,
$g(r)$ for $T \le 1307 $\,K  
(figure~\ref{fig:3}, top), 
{\it (ii)} the temperature dependence of the diffusion coefficient 
$D$ (figure~\ref{fig:4},top) and {\it (iii)} the
snapshots of the \C60 configurations at 
different temperatures (figure~\ref{fig:5}).
The incipient solid phase is characterized by
a fcc crystalline arrangement, as visible from an inspection
of the peaks' positions in the $g(r)$.

\begin{table}
\caption{Crystallization ($c$), melting ($m$) and glass ($g$) 
transition parameters. 
Melting data refer to the defective crystal with $N=1000$.
Pressures are given in MPa, temperatures in K, densities in nm$^{-3}$,
and diameters in nm; 
in the last column the packing fraction 
$\eta_{\rm x}=\pi/6\rho_{\rm x}\sigma_{\rm BH}^3$.
}
\label{tab:1}
\begin{center}
\begin{tabular}{ccccc}
 $P$ & $T_{\rm c}$  & $\rho_{\rm c}$  
& $\sigma_{\rm BH}$ & $\eta_{\rm c}$\\
\hline
 3.5       &1307        &1.124    &0.976     &0.547  \\
  40       &1464        &1.130    &0.975     &0.548  \\
 150       &1880        &1.138    &0.972     &0.547  \\
 250       &2200        &1.150    &0.970     &0.550  \\ 
$P$ & $T_{\rm m}$  & $\rho_{\rm m}$
& $\sigma_{\rm BH}$ & $\eta_{\rm m}$\\
 \hline
3.5     &1931        &1.144    &0.9716    &0.549  \\
$P$ & $T_{\rm g}$  &  $\rho_{\rm g}$
&  $\sigma_{\rm BH}$ &  $\eta_{\rm g}$\\
\hline
          3.5      &1100   &1.168    & 0.978     &0.572  \\
          40       &1170   &1.177    & 0.977     &0.575  \\
         150       &1480   &1.187    & 0.975     &0.576  \\
         250       &1700   &1.190    & 0.973     &0.574  
\end{tabular}
\end{center}
\end{table}

 The variation of $N$ from 1000 to 864 does not produce any significant
effect either on the overall cooling path, or on the location
of the turning points which marks the onset
of crystallization. 
Such an outcome  
is obtained if we evolve the smaller system 
for a longer time. In fact, under equal
time elapsed conditions the turning point with 864 particles would
fall at lower temperatures.
We interpret this result as a manifestation of 
the fact, also emerging in heating
cycles, that the 1000-particle case,
characterized by a larger simulation box and defects with respect
to the perfect fcc arrangement, 
easier allows for the internal particle rearrangements
which trigger the phase transitions.
We also note that
the first of the turning points,
signalling the onset of crystallization falls practically in 
coincidence of the crossing between the cooling path and the 
$\eta^{\rm HS}_{\rm m}=0.548$
locus introduced in equation~(\ref{eq:etamelt})
(compare figure~\ref{fig:1} and figure~\ref{fig:2});  
we shall discuss this point in the next section. 

\begin{figure*}
\begin{center}
\includegraphics[width=8.0cm,angle=-90]{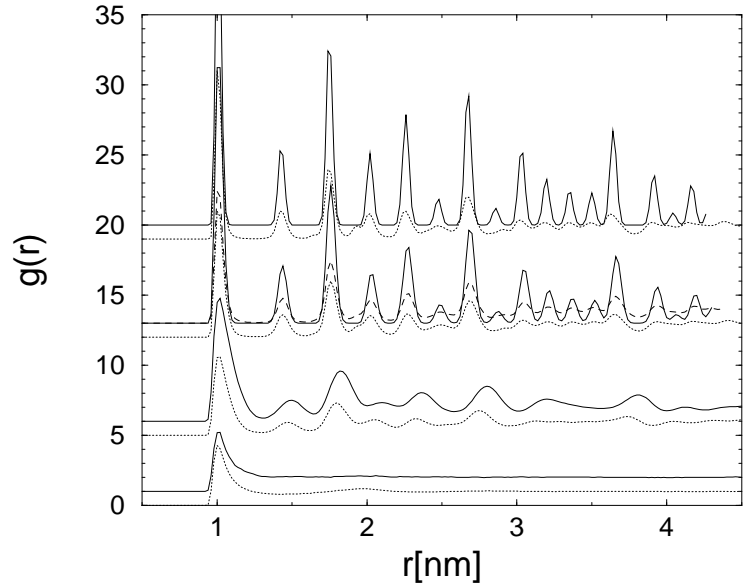}
\caption{\label{fig:6}
Evolution of $g(r)$ during the heating sequence for 
the perfect crystal with $N=864$ (full line)
at $T=598$, 1030, 2346, and 2391\,K (from top to bottom),
and for the defective crystal with $N=1000$ (dotted line) 
at $T=584$, 1037, 1931, and 1934\,K.
The dashed line is $g(r)$  at $T=1022~K$, as obtained by cooling 
the $N=864$ sample
from high temperature (1937\,K).
The comparison with the 1037\,K heating 
case enlights the smoother features
 resulting from cooling.
}\end{center}
\end{figure*}

No appreciable effect on the system behavior is
exerted by the cooling rate since 
cooling cycles with  $\Delta T=15$\,K  
lead to substantially similar results (see I).
Sensitive changes are instead associated to the value of the imposed
pressure: 
the modifications undergone by the cooling paths
as $P$ varies from 3.5 to 150 and 250\,MPa are shown
in figure~\ref{fig:1} (middle panel, see also the inset)
and figure~\ref{fig:2} (compare top and middle panels); 
the radial distribution functions for the case $P=250$\,MPa
are shown in the bottom panel of figure~\ref{fig:3} and 
numerical values of the 
crystallization parameters are reported in table~\ref{tab:1}.
It appears that the increase of the pressure 
rises dramatically the temperature where the 
onset of crystallization occurs. 
The densities attained at crystallization, and
at the end of the cooling paths, 
are close to each other and exhibit  a trend to increase
 with the pressure.
These densities
are all smaller than those of the perfect crystal heated at
the corresponding temperature, indicating
that defective crystals are formed.
 One can realize that this is so
by comparing the physical characteristics of
such crystals with those typical
of a perfect crystal of 864 particles.
 We consider to this aim crystals at room temperature
as previously obtained through the cooling from high temperatures of
samples with $N=1000$ and $N=864$, and
a perfect crystal of 864 particles under the same temperature
and pressure conditions.
As visible from figure~\ref{fig:1} and figure~\ref{fig:2},
 the final densities  of the samples
obtained from the cooling procedure are smaller than that of the
perfect 864 crystal. Such lower density values
must be attributed to the presence of voids in the crystalline
matrix formed during the cooling process.
The comparison of radial distribution functions
in figure~\ref{fig:6} enlights the smoother structure of 
cooled samples with respect to the perfect crystal.  
 At P=250\,MPa, after the onset of crystallization, 
the cooling path runs almost overlapped
 with the melting curve of \C60 \cite{costa1} 
(see figure~\ref{fig:1}, middle). 
The high pressure is thus able to force 
the system to evolve out of the metastable
region approximately along the true coexistence line.

The transition to the crystal takes place under
strong supercooling conditions,
well beyond the freezing line. indeed, when the latter is crossed,
no feature in the radial distribution function signals
the onset of any structural order typical of the solid
phase (see figure~\ref{fig:3}).
 Signatures albeit faint of such a crossing
can be recognized however in the $V$ and $H$ patterns 
which show a corresponding
tiny nonmonotonicity (see top panels
of figure~\ref{fig:2} and insets),
as well as in 
the diffusion coefficient (see figure~\ref{fig:4} and insets). 
As visible,
runs with different number of particles
exhibit the same features, ruling out
the possibility that our observations are merely a consequence
of statistical noise.

A similar situation emerges when the systems approaches
the metastable portion of the liquid-vapor binodal of \C60, calculated in 
reference~\cite{fucile} and reported in figure~\ref{fig:1}. It appears
that the cooling paths get closer to the binodal
the lower the pressure becomes.
At $P=3.5$\,MPa a crossing can be extrapolated to take place at
$T=1700$\,K and $\rho=0.95$\, nm$^{-3}$, 
a state point where $V$ and $H$ patterns
show another nonmonotonicity, magnified in the insets of 
figure~\ref{fig:2}. 
Remarkably, 
also the diffusion coefficient in figure~\ref{fig:4} is 
nonmotonic in correspondence of the estimated crossing of the binodal.
We have carried out 
very long simulation runs (up to several million time steps)
in correspondence of the crossing
 of the freezing and binodal lines,
to check whether the system is able to
 develop any signal of the incipient transitions.
 In no case, however, we could monitor  
transformations in the thermodynamic and diffusion coefficient behavior
comparable to those heralding at the crystallization temperature 
 $T=1307$\,K, as previously described. 
Several similarities emerge in comparison with
the simulation study that two of us have performed~\cite{ballone}
on a modified Lennard-Jones potential used to model
globular protein solutions~\cite{tenwolde}.
In that case, 
the transition to the crystalline arrangement takes place 
during isochoric coolings of the system 
deeply beneath the freezing line, when the system reaches
the metastable binodal line.
In the present case, the system must be cooled 
even substantially below the metastable binodal
before crystallization can start.

Starting from the two solid simulation samples  
eventually obtained through the cooling cycles
with $N=1000$ and $864$, and from a third one 
built as a perfect crystal with 864 particles,
we gradually heat all these systems 
through successive $\Delta T$=30\,K increases, 
up to the temperature at which they undergo
an abrupt transition to a fluid configuration.
Bottom panels of figures~\ref{fig:1} and~\ref{fig:2} refer
to this sequences;
it appears that the defective sample with
one thousand particles undergoes a true melting since the systems
jumps to a thermodynamic point inside the liquid
pocket of the phase diagram whereas 
the 864 particle crystal, 
either defective or perfect, is instead
substantially overheated until it
jumps directly to a vapor phase, thus exhibiting sublimation. 
The role of defects at intermediate temperatures emerges
by the comparison of
the radial distribution functions
obtained from the heating of the perfect crystal at
relatively high temperature with the 
corresponding $g(r)$ of a crystal
at an equal temperature obtained from cooling: 
in figure~\ref{fig:6} one can see that
all features are less sharp in the 
latter case, with some peaks even missing with respect
to the perfect crystal case. 
As  can be seen 
in the corresponding panels in figure~\ref{fig:2},
an evident ``hysteresis'' characterizes the heating cycles
with respect to the cooling sequences.

\subsection{Quenching and glass transition} 

Quenching of the liquid at different pressures 
is carried out for $N=1000$ and 864, starting
from an initial temperature 
$T=1950$\,K, either through a sequence of $\Delta T= 150$\,K 
decrease steps,  
or through a sudden $\Delta T$=1000\,K drop. 
Temperature variations are imposed over 20\,000 simulation
steps, followed by 10\,000 steps of free evolution.
The quenching paths are displayed in figure~\ref{fig:1} 
(top and middle panels)
while the volume and enthalpy behavior is 
visible in figure~\ref{fig:2} (top and middle panels).

\begin{figure*}
\begin{center}
\includegraphics[width=8.4cm,angle=-90]{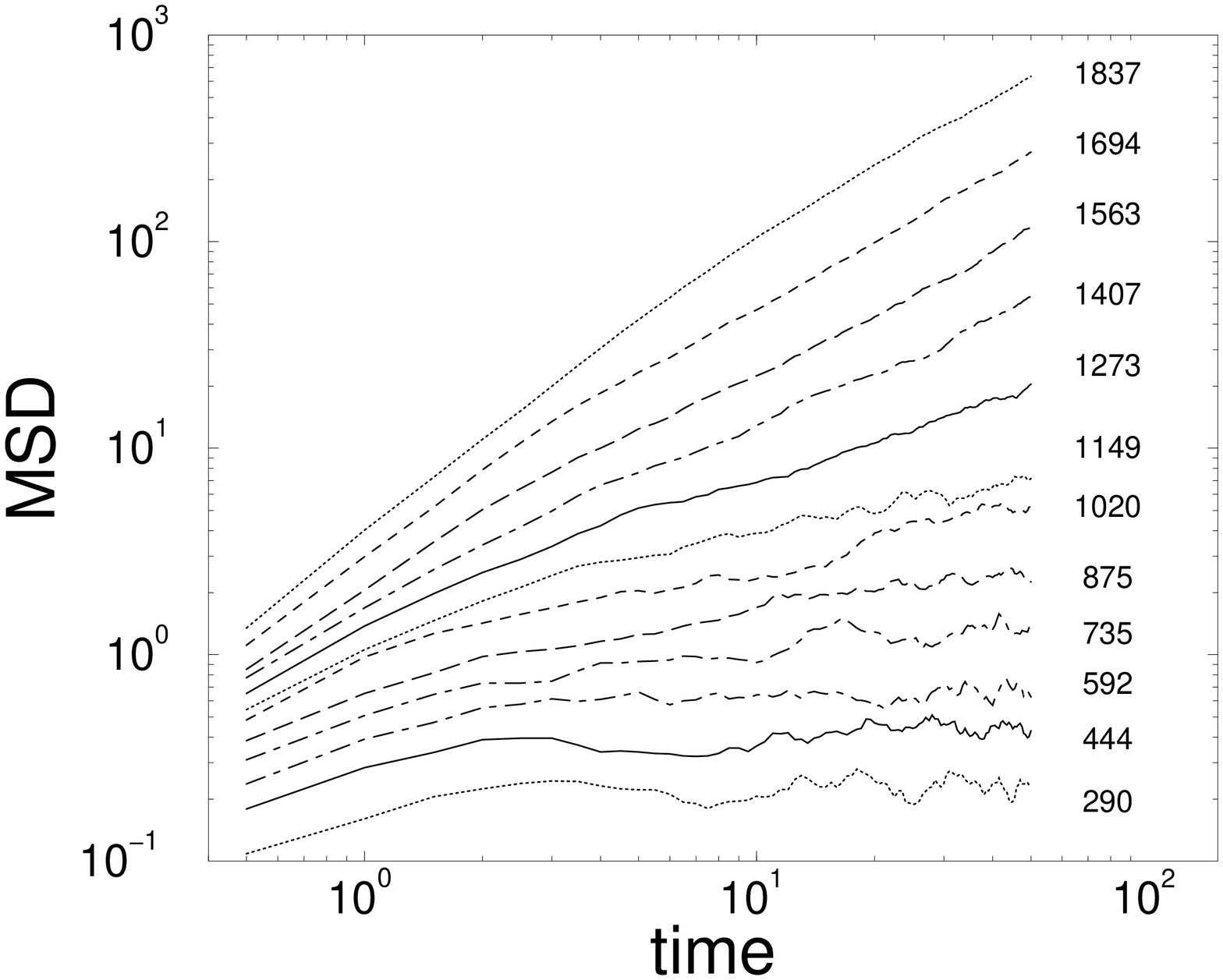}\\
\includegraphics[width=8.5cm,angle=-90]{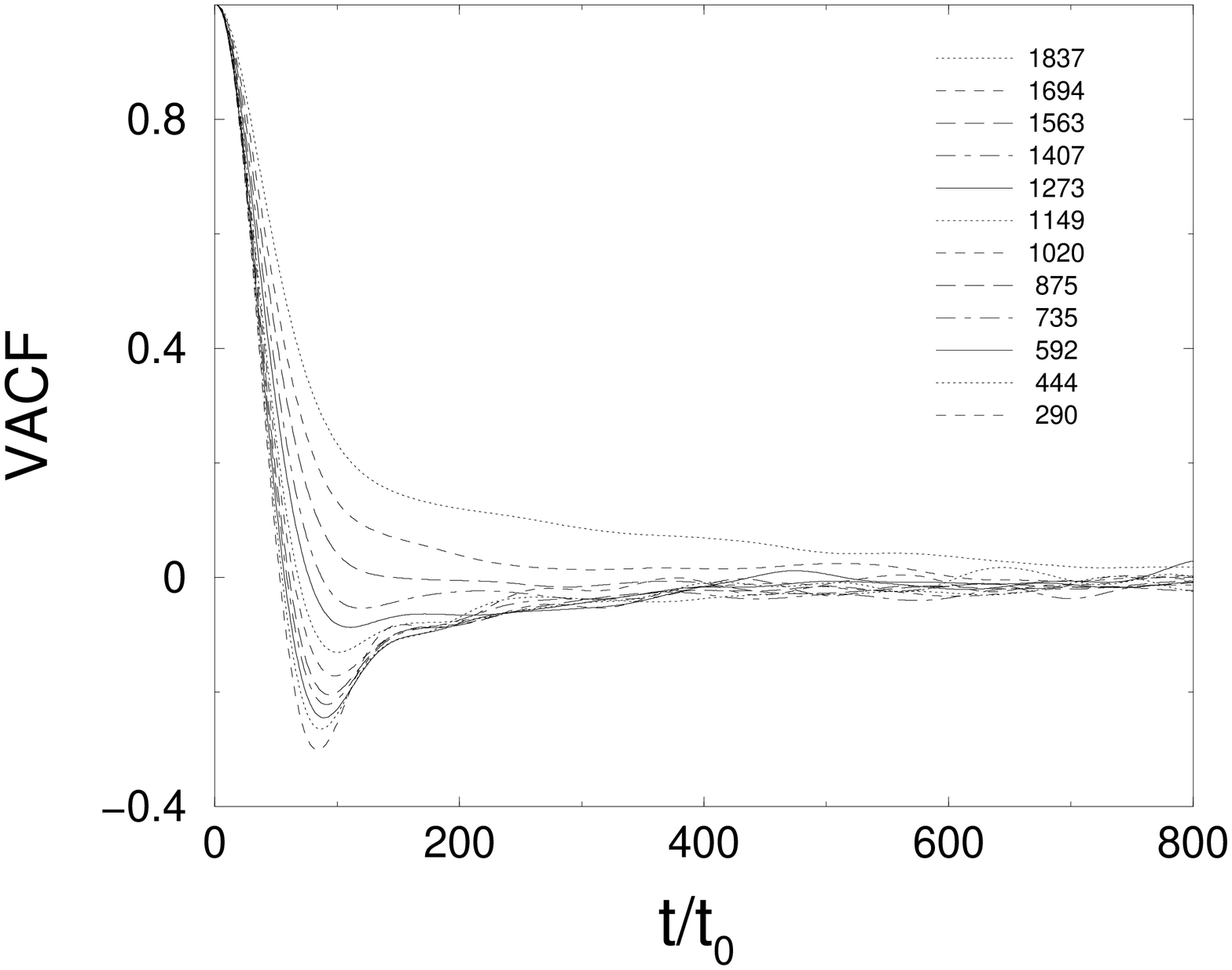}
\caption{\label{fig:7}
{\it Top:} Mean square displacement (in \AA/s units) in the $N=864$ sample
as a function of the simulation time (in picoseconds) 
during  the quenching sequence at $P=3.5$\,MPa.
{\it Bottom:} Evolution of the velocity autocorrelation function 
during the same sequence ($t_0$=time step).} 
\end{center}
\end{figure*}

\begin{figure*}
\begin{center}
\includegraphics[width=9.0cm]{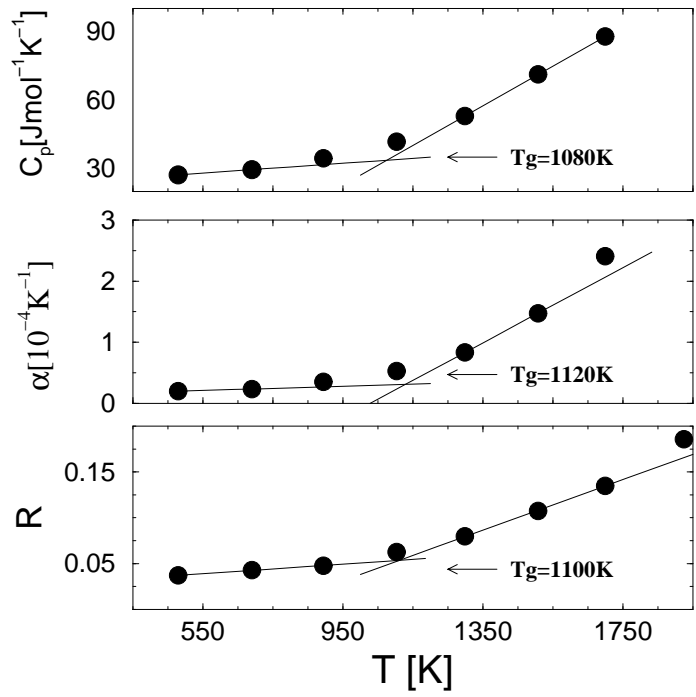}
\bigskip
\caption{\label{fig:8}
Heat capacity (top), thermal expansion (middle),
and Wendht-Abraham ratio (bottom) as functions of the temperature during 
the quenching sequence of the $N=864$ sample at $P=3.5$\,MPa.
}\end{center}
\end{figure*}

The onset of a glassy phase during both quenching procedures 
is visually documented in the snapshots shown in figure~\ref{fig:5} and
is proved by several evidences on thermodynamic,
structural and dynamical quantities.
We first observe that
the $V$ and $H$ patterns in figure~\ref{fig:2}
show a glass branch running above the crystallization line; moreover,
at variance with what we have observed during the slow cooling, 
the radial distibution functions in figure~\ref{fig:3} do not develop 
the peak structure of the fcc arrangement at low temperatures; rather,
down to $\simeq 300$\,K they exhibit the twofold
structure of the second peak typical of the glass.
Such a structural evidence has a counterpart in the diffusion
coefficient behavior which, as visible in figure~\ref{fig:4},
does not exhibit the drop associated to the onset of
crystallization process, but manifests only a change of slope.
On the other hand, 
the mean square displacements
show a plateau for temperatures $T \le 1100$\,K,
signalling structural arrest, 
also confirmed by the shape variations of  
the velocity autocorrelation function with 
the temperature (see figure~\ref{fig:7}).
Changes of slope as the temperature decrease,
 typically associated to the glass
transition, occur 
also in $V$ and $H$ (figure~\ref{fig:2}), as well as 
in the thermal expansivity $\alpha$,
the specific heat $C_{\rm p}$, 
and the Wendt-Abraham ratio $R$ of the first peak to the first
minimum height in $g(r)$ (figure~\ref{fig:8}). 

\begin{figure*}
\begin{center}
\includegraphics[width=9.0cm,angle=-90]{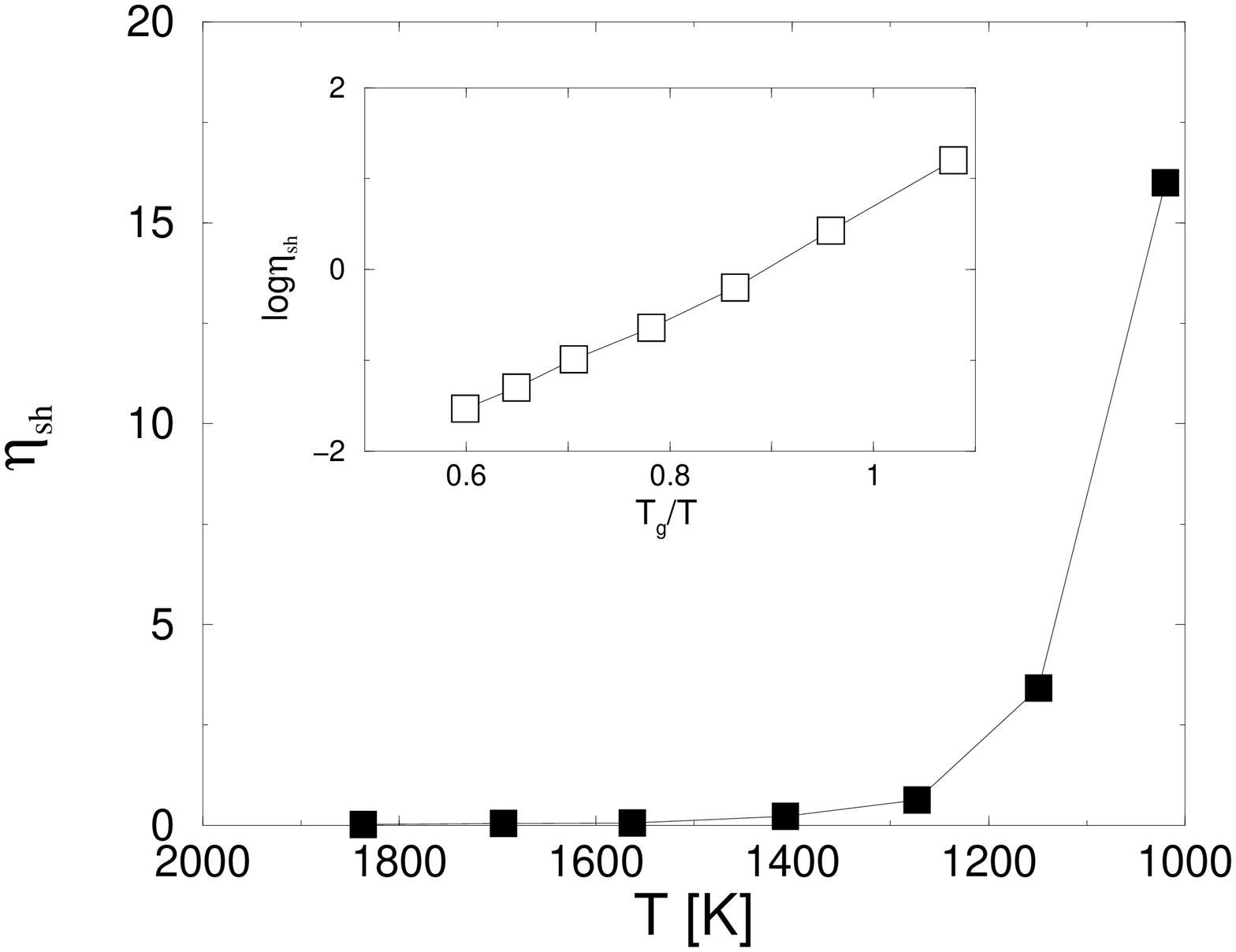}
\caption{\label{fig:9}
Shear viscosity vs temperature during a quenching sequence of the 
$N=864$ sample at $P=3.5$\,MPa.
The inset shows the Arrhenius behaviour in the approach 
to the glass temperature.} 
\end{center}
\end{figure*}

Evidence of the glass formation also comes from the shear viscosity
which we have calculated via the Green-Kubo relation~\cite{hansen}: 
\begin{equation}\label{eq:visc1}
\eta_{\rm sh}=\frac{1}{Vk_{\rm B}T}\int_0^\infty 
\langle \sigma^{xy}(0)\sigma^{xy}(t)\rangle  \diff t \,,
\end{equation}
where
\begin{equation}\label{eq:visc2}
\sigma^{xy}=\sum_{i=1}^N \left[ 
m_iv_i^xv_i^y+\frac{1}{2}\sum_{j\ne i}x_{ij}f_y(r_{ij}) \right]
\end{equation}
is the off-diagonal component
of the stress tensor,
$v_i^x$ and $x_{ij}$ represent, respectively, 
the $x$ component of the velocity and separation distance, $r_{ij}$,
between molecules' center-of-mass $i$ and $j$, and $f_y$
is the $y$ component of the force exerted on atom $i$ by atom $j$.
As can be appreciated in figure~\ref{fig:9}
$\eta_{\rm sh}$ exhibits a marked increase across the 
glass transition region and 
displays an Arrhenius behavior as a function of $T$ (see inset), 
indicating that the glass
formed is a ``strong'' one~\cite{angell};
the orientational glass of \C60
is similarly ``strong''~\cite{gugenberger,matsuo}.

 The glass transition parameters are obtained from   
the intersection of the two (extrapolated) branches
 in the  $V$,
 $H$, $C_{\rm P}$, $\alpha$ and $R$ patterns as functions of the temperature
(see reference~\cite{angell} for details):  
we estimate at $P=3.5$\,MPa
 $T_{\rm g} \simeq 1100$\,K 
and $\rho_{\rm g} \simeq 1.168$\,nm$^{-3}$.

As visible from figure~\ref{fig:2} a glass 
transition is observed also at pressures higher than $3.5$\,MPa,
at the cost to increase
the quenching rate
with respect to $1.5 \times 10^{12}$\,K/s adopted at 3.5\,MPa.
For instance, at $P=250$\,MPa no glassy phase is
 obtained if $\Delta T \le 300$\,K over 20\,000 time steps.
Glass transition temperatures and densities
at different pressures are reported in table~\ref{tab:1}.
It is immediate to verify 
that, similarly to what observed
with the crystallization transitions, 
the glass transitions (at different pressures)
occur for states falling at the crossing of the quenching
paths with the $\rho(T)$ locus defined by equation~(\ref{eq:etaglass}),
where $\eta_{\rm g}$= $\eta_{\rm g}^{\rm HS}$.
We shall comment on these evidences
in the next section.  

The quenching rates we adopt in our simulations 
are much higher than those presently achieved in experiments (typically
$10^8$\,K/s). We have however preliminary simulative evidences~\cite{miscela}
that in equimolar Girifalco \C60/C$_{70}$ and \C60/C$_{96}$ mixtures
crystallization does not occur even upon
cooling rate one or two order of magnitude lower
than those adopted here; 
the \C60/C$_{70}$ mixture, for instance, 
remains liquid during the cooling procedure 
down to 1100\,K, a temperature lower than that at which amorphization
of pure \C60 fullerite occurs~\cite{fischer} (see also note \cite{nota}).
These findings make it plausible that 
a glassy phase might be formed experimentally at least 
from mixed fullerene systems.

\section{Discussion and conclusions}

\begin{figure*}
\begin{center}
\hspace*{8pt}\includegraphics[width=6.6cm,angle=-90]{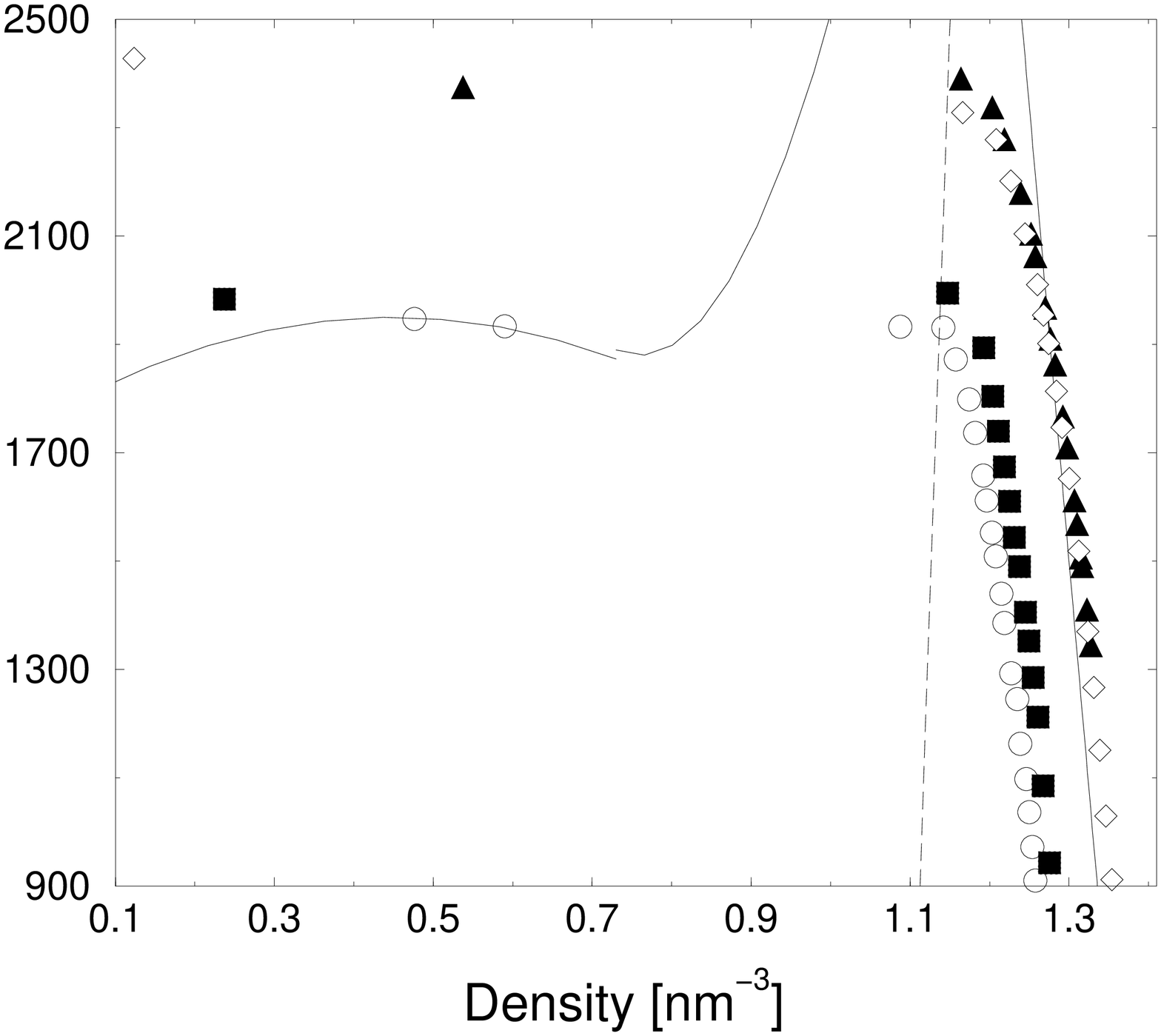}
\caption{\label{fig:1}
{\it Top:} Continuous lines: phase diagram of \C60~\cite{costa1}
 with superimposed  the 
$\rho_{\rm m}(T)$ [long-dashed line, equation~(\ref{eq:etamelt})] and
$\rho_{\rm g}(T)$ [short-dashed line, equation~(\ref{eq:etaglass})]
loci introduced in this work.
 Circles and squares: cooling with 1000 and 864 particles at $P=3.5$\,MPa,
 respectively. 
 Inset: Helmholtz free energy vs $\rho$  at $T=2100$\,K~\cite{costa1}. 
{\it Middle:} Effect of increasing pressure on the cooling and quenching paths.
 Full circles, squares and triangles: cooling paths with $N=1000$ at 
 pressure $P=3.5$, 150 and 250\,MPa, respectively; 
 open circles, squares and  triangles: quenching with $N=1000$ at $P=3.5$,
 150 and 250\,MPa; diamonds: quenching at 3.5\,MPa with 864 particles.
The dot-dashed line indicates the metastable portion of the 
binodal~\cite{fucile}.
 Inset: crystallization density as a function of the pressure and
 corresponding temperatures.
 {\it Bottom:} Heating paths of the defective and perfect crystals.
 Circles and squares: heating of 
the defective crystal with $N=1000$ and
 864, respectively; diamonds and triangles: 
 heating of the perfect crystal
 with $N=864$ at $P=3.5$\,MPa and 12\,MPa, respectively.}
\end{center}
\end{figure*}

The effective packing at crystallization
for different pressures have values 
close to the packing fraction of hard spheres at melting
$\eta^{\rm HS}_{\rm m}$=0.545, as reported in table~\ref{tab:1}.
As a matter of fact, it thus apperas
that  the $\eta_{\rm m}(T)$ locus introduced 
in equation~(\ref{eq:etamelt}) 
acts as an interpolation curve among
the transition points
determined {\it a posteriori}, 
and can be used to forecast 
crystallization temperatures and densities at pressures other than
those here considered. 
A similar evidence is found for
the effective packing at the glass transition 
  $\eta_{\rm g}(T)$ of equation~(\ref{eq:etaglass})
also reported in the table,
 that is quite
close to the hard sphere glass transition value,
$\eta_{\rm g}^{\rm HS}=0.58$~\cite{woodcock}. 
As for the onset of
crystallization and glass transitions,
 an obvious deduction 
 is that the Girifalco model
follows rather closely a hard-sphere-like behavior.
Our evidences, however, allow us to get a deeper insight
into the role of the 
$\rho_{\rm m}(T)$ and $\rho_{\rm g}(T)$ loci.
Indeed, we have found that 
when supercooling is pushed beyond
$\rho_{\rm m}(T)$ the system invariably crystallizes;
by contrast, a state close to, but on the left of 
$\rho_{\rm m}(T)$,
remains in the supercooled phase even over the longest 
 simulation run we could perform, namely 12 million simulation steps,
corresponding to 60\,ns.
It thus appears that the $\rho_{\rm m}(T)$
locus approximately corresponds to 
an instability boundary of the metastable region.
This observation  also emerges from Monte Carlo calculations
of the free energy of the Girifalco model
at $T=2100$\,K~\cite{costa} (see inset of figure~\ref{fig:1}); 
in fact,
$\rho_{\rm m }$(2100\,K) coincides with the point where the
free energy shows an inflection point associated with a sudden drop of 
the pressure of the simulation sample,
i.e. with a mechanical instability of the system. 
A similar property might characterize the free 
energy branch generated by approaching  
the metastable region from the solid side and
since the interval separating the 
two branches in figure~\ref{fig:1} is quite narrow, it is conceivable 
that our simulation strategy is unable to
discriminate between the densities at the crystallization and melting
instabilities. 
It is interesting to observe that in Molecular Dynamics simulations of the
Lennard-Jones potential, crystallization 
occurs much beyond the melting line, deeply
inside the solid region~\cite{nose}. Indeed
we have carried out MD simulations 
of the same potential and 
verified that states located inside 
the metastable fluid-solid region, 
do not undergo crystallization even
after 35 million time steps ($\simeq$ 180\,ns). 

 Our results fot the glass transition
positively agree with the 
 predictions made by Foffi and coworkers~\cite{foffi} 
on the glass transition in protein solutions modeled through 
the attractive hard-core Yukawa (HCY) potential.
These authors find that 
for inverse decay length {\it z} of the Yukawian term
spanning from 5 to 60 times the reciprocal of the particle diameter,
the glass transition line is an almost vertical locus 
with a $T= \infty$ asymptote $\eta=\eta_{\rm g}^{\rm HS}=0.58$.
On the other hand, as discussed  in~\cite{hagen2}, 
the physical properties of 
the Girifalco model can be reasonably reproduced by a
Yukawa potential with $z \simeq 4$ (though with moderate differences
in the phase diagram), close indeed to the lowest value
investigated in reference~\cite{foffi}.
A  qualitative agreement thus emerges between the prediction
of the glass transition in the two different systems. 
In the HCY model the formation 
of ``repulsive'' and ``attractive'' glasses depends
on the value of $z$~\cite{foffi};
since at $z$ as low as five
 only the repulsive glass would be formed,
 we identify our \C60 ``positional'' glass as a repulsive one.

In conclusion, we have documented that 
effective hard-core exclusion effects
 play an important role for the determination of the glass transition 
in a short-range potential
well suited to model various fullerenes.
Mode Coupling Theory calculations  
for the onset of the glass transition in a 
short-range model of globular protein solutions lead to similar 
conclusions at least for relatively low values of the decay potential
parameters.  
In this context,
it could be worth to
carry out specific Mode Coupling calculations for the 
present model fullerene
in order to compare the structural arrest line 
thereby predicted with our results for the glass line
based on simulations 
 and $\rho_{\rm g}(T)$ locus predictions.
Moreover,
colloidal suspensions and colloid-polymer mixtures
show a behavior characterized by  several similarities
with the observations reported in this work~\cite{louis}. 
We argue that the basic equation~(\ref{eq:etaglass})  
might hold to a high accuracy also for other interaction potentials,
and hence it could provide a simple framework for a qualitative
but immediate prediction of the glass line in a variety of model systems. 

\acknowledgments
This work has been done in the framework of the Marie Curie Network
on Dynamical Arrest of Soft Matter and Colloids,
Contract Nr MRTN-CT-2003-504712.

\newpage


\begin{figure*}
\begin{center}
\hspace*{2pt}\includegraphics[width=5.7cm,angle=-90]{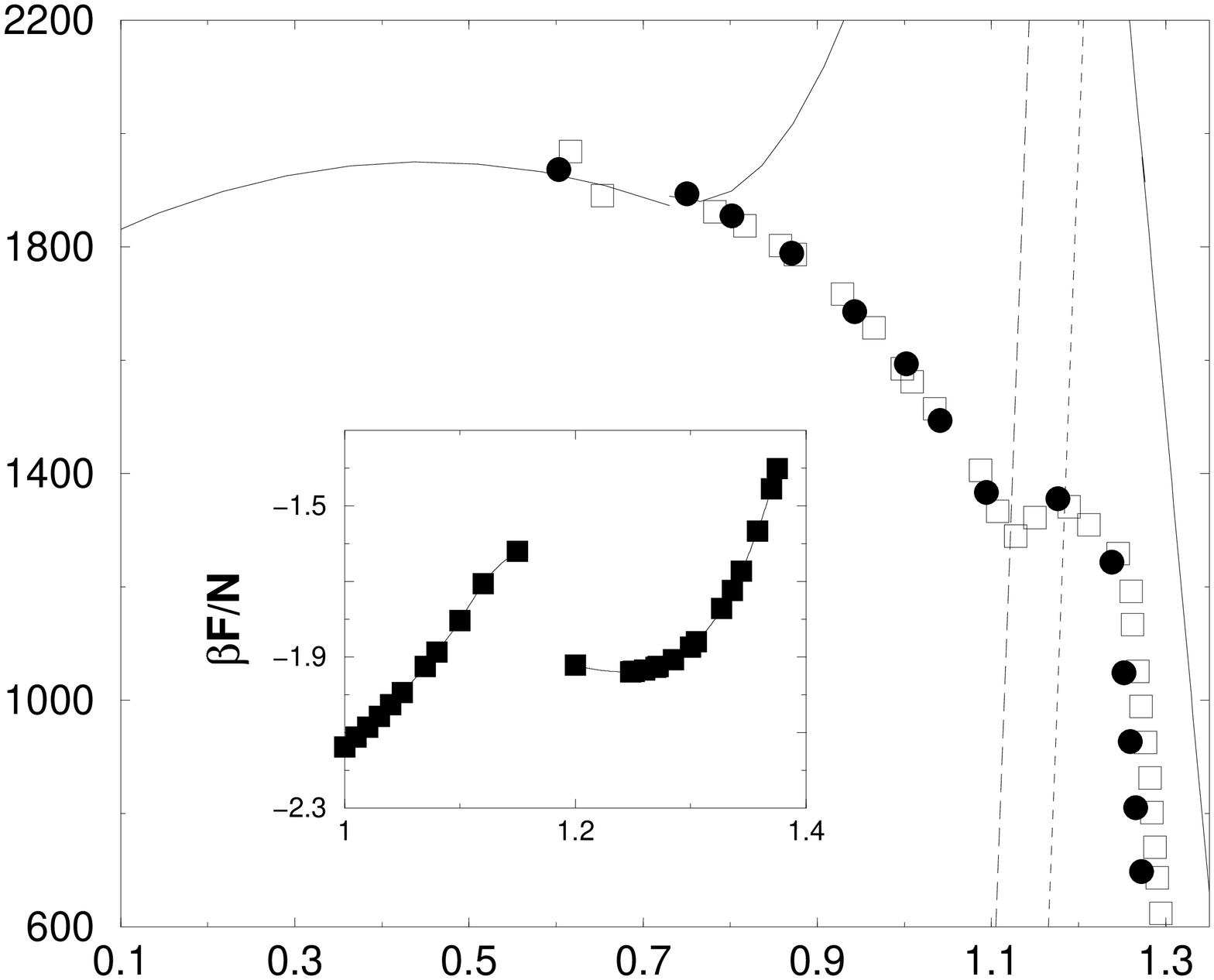}
\end{center}
\end{figure*}
\begin{figure*}
\begin{center}
\includegraphics[width=5.8cm,angle=-90]{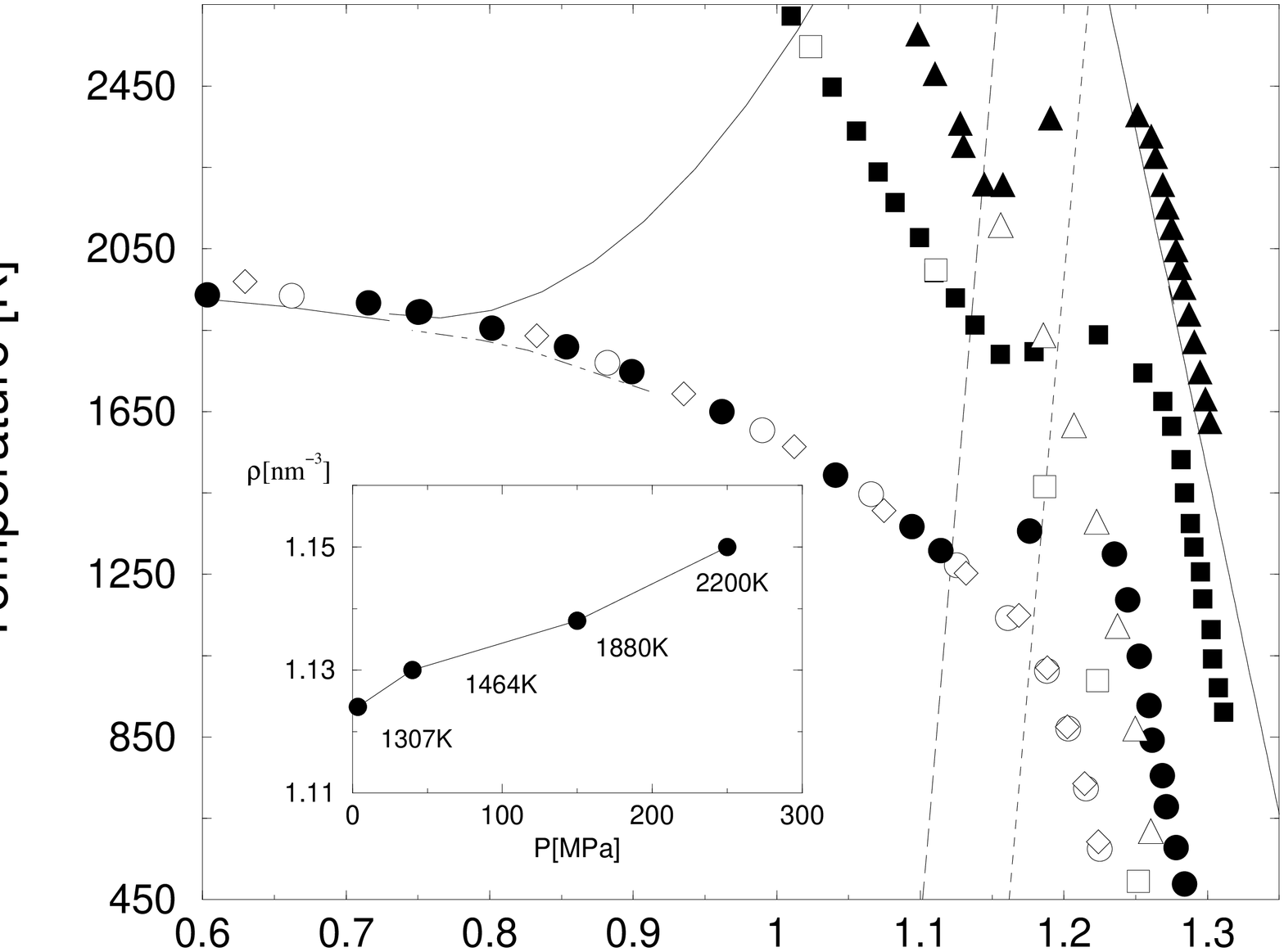}\\
\end{center}
\end{figure*}

\begin{figure*}
\begin{center}
\includegraphics[width=6.0cm,angle=-90]{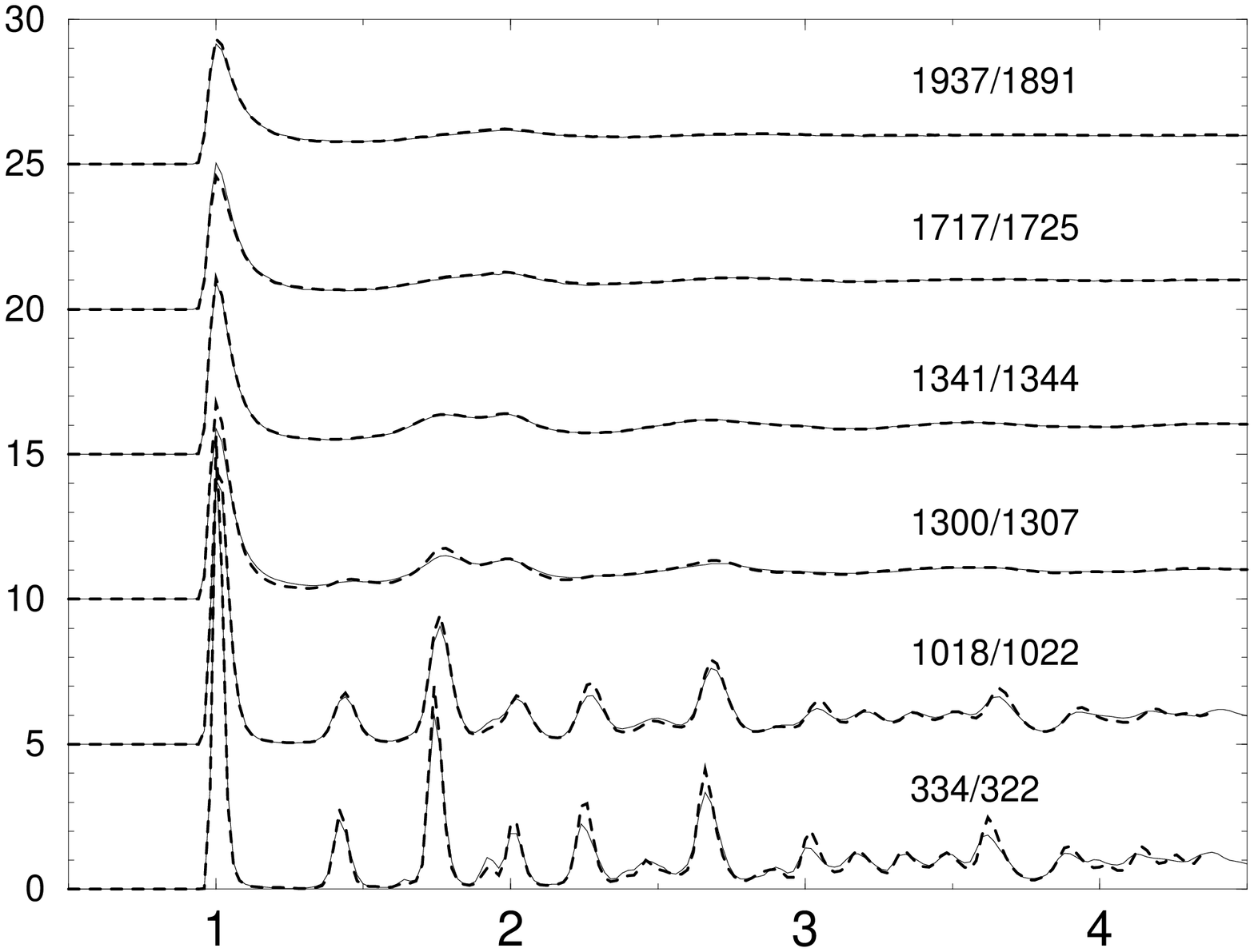}
\end{center}
\end{figure*}
\begin{figure*}
\begin{center}
\includegraphics[width=6.5cm,angle=-90]{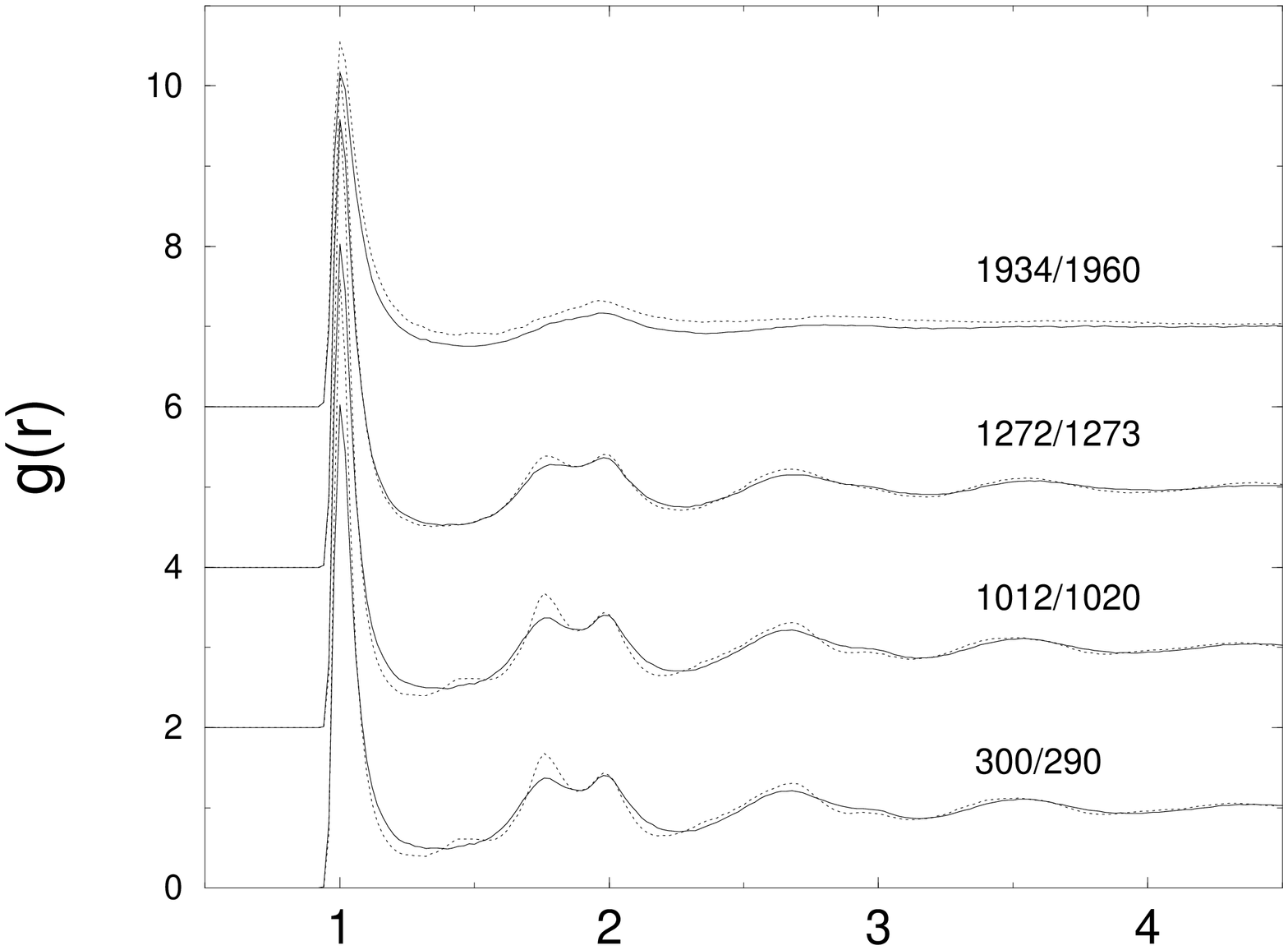}
\end{center}
\end{figure*}

\begin{thebibliography}{99}
\bibitem{nota}
Fullerite (crystalline \C60) heated at $T >1200 $\,K, 
mostly transforms in amorphous carbon due to the
disruption of the \C60 cages (see ref~\cite{fischer}). The latter, however, 
are predicted to be stable up to 3500-4000\,K~\cite{xu}. 
Solid state effects through molecule collision~\cite{sundar},
but also and alternatively residual impurities
trapped in the crystalline matrix~\cite{noi}, 
might trigger the cage instability.
\end{thebibliography}
\end{document}